\documentclass[3p]{earticle-nodate}

\PassOptionsToPackage{hyphens}{url}
\usepackage{graphicx}
\usepackage{dcolumn}
\usepackage{bm}
\usepackage[T1]{fontenc}
\usepackage[utf8]{inputenc}
\usepackage{amssymb}
\usepackage{amsmath}

\usepackage[dvipsnames]{xcolor}
\usepackage{placeins}
\usepackage{listings}
\usepackage{setspace}
\usepackage[normalem]{ulem}

\usepackage[hyphens]{url}
\usepackage{hyperref}
\usepackage[hyphenbreaks]{breakurl}

\usepackage{xr}
\usepackage{tabularx}
\usepackage{lmodern}

\biboptions{super}

\newcommand{\VEs}{\text{VEs}}
\newcommand{\VEd}{\text{VEd}}
\newcommand{\VEc}{\text{VEc}}
\newcommand{\VEi}{\text{VEi}}

\newcommand{\VEf}{\text{VEf}}
\newcommand{\VEh}{\text{VEh}}

\begin{document}

\abstracttitle{Summary}

\begin{frontmatter}

\title{Simulating transmission scenarios of the Delta variant of SARS-CoV-2 in Australia}

\author{Sheryl L. Chang$^{1}$, Oliver M. Cliff$^{1,2}$, Cameron Zachreson$^{1,3}$,
Mikhail Prokopenko$^{1,4}$}
\address{$^{1}$  Centre for Complex Systems, Faculty of Engineering, The University of Sydney, Sydney, NSW 2006, Australia\\
$^{2}$  School of Physics, The University of Sydney, Sydney, NSW 2006, Australia\\
$^{3}$  School of Computing and Information Systems, The University of Melbourne, Parkville, VIC 3052, Australia\\
$^{4}$  Sydney Institute for Infectious Diseases, The University of Sydney, Westmead, NSW 2145, Australia\\ 
  \ \\
Correspondence to:\\ Prof Mikhail Prokopenko, Centre for Complex Systems\\ Faculty of Engineering, University of Sydney, Sydney, NSW 2006, Australia\\ \textbf{mikhail.prokopenko@sydney.edu.au} 
}


\begin{abstract}
An outbreak of the Delta (B.1.617.2) variant of SARS-CoV-2 that began around mid-June 2021 in Sydney, Australia, quickly developed into a nation-wide epidemic. The ongoing epidemic is of major concern as the Delta variant is more infectious than previous variants that circulated in Australia in 2020. Using a re-calibrated agent-based model, we explored a feasible range of non-pharmaceutical interventions, including case isolation, home quarantine, school closures, and stay-at-home restrictions (i.e., ``social distancing''). Our modelling indicated that the levels of reduced interactions in workplaces and across communities attained in Sydney and other parts of the nation were inadequate for controlling the outbreak. A counter-factual analysis suggested that if 70\% of the population followed tight stay-at-home restrictions, then at least 45 days would have been needed for  new daily cases to fall from their peak to below ten per day. Our model predicted that, under a progressive vaccination rollout, if 40-50\% of the Australian population follow stay-at-home restrictions, the incidence will peak by mid-October 2021: the peak in incidence across the nation was indeed observed in mid-October. We also quantified an expected burden on the healthcare system and potential fatalities across Australia.
\end{abstract}

\end{frontmatter}

\section*{Introduction}


Strict mitigation and suppression measures eliminated local transmission of SARS-CoV-2 during the initial pandemic wave in Australia (March--June 2020; peaked around 500 cases per day, i.e., around 20 daily cases per million)~\citep{chang2020modelling}, as well as a second wave that developed in the south-eastern state of Victoria (June--September 2020; peaked around 700 cases per day, i.e., around 30 daily cases per million)~\citep{blakely2020probability,zachreson2021risk}\footnote{In describing a ``wave'' we follow the definition based on two key features: (i) an epidemic wave comprises upward and/or downward periods; and (ii) the increase during an upward period, as well as the decrease during a downward period, must be substantial over a period of time~\citep{zhang2021second}.}.  
Several subsequent outbreaks were also detected and managed quickly and efficiently by contact tracing and local lockdowns, e.g., a cluster in the Northern Beaches Council of Sydney, New South Wales~(NSW) totalled 217 cases and was controlled in 32 days by locking down only the immediately affected suburbs (December 2020--January 2021)~\citep{covid19data}. Overall, successful pandemic response was facilitated by effective travel restrictions and stringent stay-at-home restrictions (i.e., ``social distancing''), underpinned by a high-intensity disease surveillance~\citep{rockett2020revealing,scott2020modelling,milne2021modelling,lokuge2021exit,wilson2021estimating}.  

Unfortunately, the situation changed in mid-June 2021, when a highly transmissible variant of concern, B.1.617.2 (Delta), was detected. The first infection was recorded on June 16 in Sydney, and quickly spread through the Greater Sydney area. Within ten days, there were more than 100 locally acquired cumulative cases, triggering stay-at-home (social distancing) restrictions imposed in Greater Sydney and nearby areas~\citep{nsw-jul-2021}. By July 9 (23 days later), the locally acquired cases reached 439 in total~\citep{covid19data}, and a tighter lockdown was announced~\citep{nsw-jul-2021}. Further restrictions and business shut-downs, including construction and retail industries, were announced on 17 July~\citep{lockdown-jul-17-2021}. By then, the risk of a prolonged lockdown had become apparent~\citep{theage}, with the epidemic spreading to the other states and territories, most notably Victoria (VIC) and the Australian Capital Territory (ACT). The incidence peaked, around 2,750 daily cases, i.e., around 100 daily cases per million, only by mid-October 2021, and stabilised in November within the range between 1,200 and 1,600 daily cases, i.e., between 45 and 65 daily cases per million~\citep{covid19data}, before a new surge of infections in December 2021 due to the Omicron variant (B.1.1.529).

The difficulty of controlling the third epidemic wave (June--November 2021) is attributed to a high transmissibility of the B.1.617.2 (Delta) variant, which is known to increase the risk of household transmission by approximately 60\% in comparison to the B.1.1.7 (Alpha) variant~\citep{phe-11-june-2021}. This transmissibility was compounded by the initially low rate of vaccination in Australia, with around 6\% of the adult population double vaccinated before the Sydney outbreak and only 7.92\% of adult Australians double vaccinated by the end of June 2021~\citep{aus-01-jul-2021}, with this fraction increasing to 67.24\% by 15 October 2021 and 83.01\% by 13 November 2021~\citep{aus--2021}.


Several additional factors make the Sydney outbreak and the third pandemic wave in Australia (June--November 2021) an important case study, in which the system complexity and the search space formed by possible interventions can be reduced.  Because previous pandemic waves were eliminated in Australia, the Delta variant has not been competing with other variants. Secondly, the level of acquired immunity to SARS-CoV-2 in the Australian population was low at the onset of the outbreak, given that (a) the pre-existing natural immunity was limited by cumulative confirmed cases of around 0.12\%, and (b) immunity acquired due to vaccination did not extend beyond 6\% of the adult population.  Furthermore, the school winter break in NSW (28 June -- 9 July) coincided with the period of social distancing restrictions announced on 26 June, with school premises remaining mostly closed beyond 9 July. Thus, the epidemic suppression policy of school closures is not a free variable, further reducing the search space of available control measures. 


This study addresses several important questions. 
Firstly, we investigate a feasible range of key non-pharmaceutical interventions (NPIs): case isolation, home quarantine, school closures and social distancing, available to control  virus transmission within a population with a low immunity.  
Social distancing (SD) is interpreted and modelled in a broad sense of comprehensive stay-at-home restrictions, comprising several specific behavioural changes that reduce the intensity of interactions among individuals (and hence the virus transmission probability), including physical distancing, mobility reduction, mask wearing, and so on.
Our primary focus is a ``retrodictive'' estimation of the average (unknown) SD level under which the modelled transmission and suppression dynamics can be best matched to the observed incidence data.  An identification of the SD level helps to distinguish and evaluate the distinct and time-varying impacts of NPIs and vaccination campaigns.

Secondly, in a counter-factual mode, we quantify under what conditions the initial outbreak could have been suppressed, aiming to clarify the extent of required NPIs during an early outbreak phase with low vaccination coverage, in comparison to previous pandemic control measures successfully deployed in Australia. This analysis highlights the challenges associated with imposing very tight restrictions which would be required to suppress the high transmissible Delta variant.

Finally, we offer and validate a projection for the peak of case incidence across the nation, formed in response to a progressive vaccination campaign rolling out concurrently with the strict lockdown measures adopted in NSW, VIC and ACT. In doing so, we predict the expected hospitalisations, intensive care unit (ICU) demand, and potential fatalities across Australia. Importantly, this analysis shows that a 10\% increase in the average SD level reduces the clinical burden approximately threefold, and the potential fatalities approximately twofold.

\section*{Methods}

We utilised an agent-based model~(ABM) for transmission and control of COVID-19 in Australia that has been developed in our previous work~\citep{chang2020modelling,zachreson2021how} and implemented within a large-scale software simulator (AMTraC-19). The model was cross-validated with genomic surveillance data~\citep{rockett2020revealing}, and contributed to policy recommendations on social distancing that were broadly adopted by the World Health Organisation~\citep{world2020combined}. The model separately simulates each individual as an agent within a surrogate population composed of about 23.4 million software agents. These agents are stochastically generated to match attributes of anonymous individuals (in terms of age, residence, gender, workplace, susceptibility and immunity to diseases), informed by data from the Australian Census and the Australian Curriculum, Assessment and Reporting Authority.  In addition, the simulation follows the known commuting patterns between the places of residence and work/study~\citep{cliff2018nvestigating,zachreson2018urbanization,fair2019creating}. Different contact rates specified within diverse social contexts (e.g., households, neighbourhoods, communities, and work/study environments) explicitly represent heterogeneous demographic and epidemic conditions (see Supplementary Material: Agent-based model). The model has previously been calibrated to produce characteristics of the COVID-19 pandemic corresponding to the ancestral lineage of SARS-CoV-2~\citep{chang2020modelling,zachreson2021how}, using actual case data from the first and second waves in Australia, and re-calibrated for B.1.617.2 (Delta) variant using incidence data of the Sydney outbreak (see Supplementary Material: Model calibration). 

Each epidemic scenario is simulated by updating agents’ states in discrete time. In this work we start from an initial distribution of infection, seeded by imported cases generated by the incoming international air traffic in Sydney's international airport (using data from the Australian Bureau of Infrastructure, Transport and Regional Economics)~\citep{cliff2018nvestigating,zachreson2018urbanization}.  
At each time step during the seeding phase, this process probabilistically generates new infections within a 50 km radius of the airport (covering the area within Greater Sydney's boundaries), in proportion to the average daily number of incoming passengers (using a binomial distribution and data from the Australian Bureau of Infrastructure, Transport and Regional Economics)~\citep{cliff2018nvestigating}.  

A specific outbreak, originated in proximity to the airport, is traced over time by simulating the agents interactions within their social contexts, computed in 12-hour cycles (``day'' and ``night'').  Once the outbreak size (cumulative incidence) exceeds a pre-defined threshold (e.g., 20 detected cases), the travel restrictions (TR) are imposed by the scenario, so that the rest of infections are driven by purely local transmissions, while no more overseas acquired cases are allowed (presumed to be in effective quarantine). Case-targeted non-pharmaceutical interventions (CTNPIs), such as case isolation (CI) and home quarantine (HQ), are applied from the outset. A scenario develops under some partial mass-vaccination coverage, implemented as either a progressive rollout, or a limited pre-pandemic coverage, as described in Supplementary Material: Vaccination modelling.

The outbreak-growth phase can then be interrupted by another, ``suppression'', threshold (e.g., 100 or 400 cumulative detected cases) which triggers a set of general NPIs, such as social distancing (SD) and school closures (SC). Every intervention is specified via a macro-distancing level of compliance (i.e., $SD = 0.8$ means 80\% of agents are socially distancing), and a set of micro-distancing parameters (quantifying context-specific interaction strengths, e.g., moderate or tight restrictions) that indicate the level of social distancing within a specific social context (households, communities, workplaces, etc.). For instance, for those agents that are compliant, contacts (and thus likelihood of infection) can be reduced during a lockdown to $SD_w = 0.1$ within workplaces and $SD_c = 0.25$ within communities, whilst maintaining contacts $SD_h = 1.0$ within households. To re-iterate, ``social distancing'' modelled in this study comprises a range of restrictions that reduce the intensity of interactions among individuals, including  mask wearing, physical distancing by several metres, mobility, and so on. We do not estimate a relative importance of these specific NPI approaches, each of which separately contributes to reducing SARS-CoV-2 transmission~\citep{li2021face,tang2021filtration,jones2020two,zhang2021non,macintyre2021use,trauer2021understanding}, focusing instead on a differentiation between the effects of NPIs and vaccination campaigns.

\section*{Results}

\begin{figure}[t]
    \centering
    \includegraphics[clip, trim=4.2cm 5.4cm 4.2cm 5.9cm, width=\textwidth]{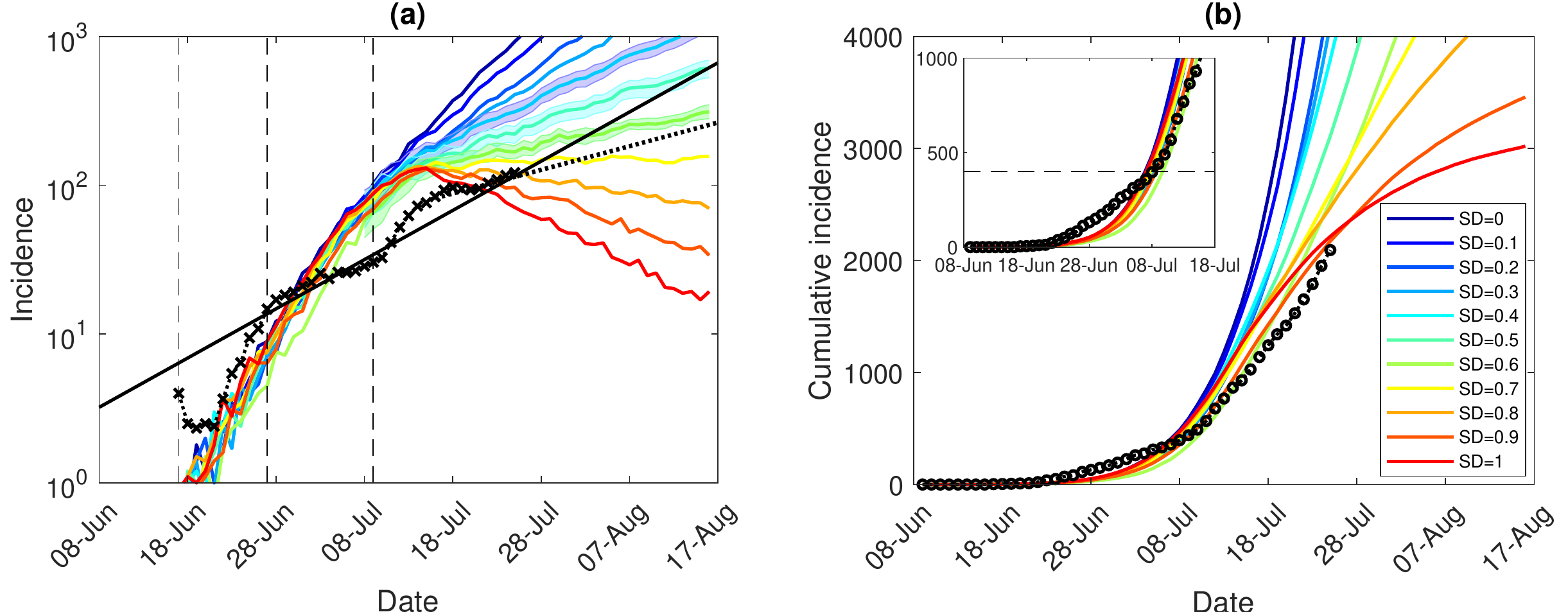} 	
    \caption{\textbf{Moderate restrictions (NSW; progressive vaccination rollout; suppression threshold: 400 cases)}: a comparison between simulation scenarios and actual epidemic curves, under moderate interaction strengths ($CI_c = CI_w = 0.25$, $HQ_c = HQ_w = 0.25$, $SD_c = 0.25$, $SC = 0.5$). A moving average  of the actual time series up to 25 July for (a) (log-scale) incidence (crosses), and (b) cumulative incidence (circles); with an exponential fit of the incidence's moving average (black solid: $\beta_{II}$, and black dotted: $\beta_{III}$). Vertical dashed marks align the simulated days with the outbreak start (17 June, day 9), initial restrictions (27 June, day 19), and tighter lockdown (9 July, day 31).  Traces corresponding to each social distancing (SD) compliance level are shown as average over 10 runs (coloured profiles for SD varying in increments of 10\%, i.e., between $SD = 0.0$ and $SD = 1.0$). 95\% confidence intervals for incidence profiles, for $SD \in \{0.4, 0.5, 0.6\}$, are shown as shaded areas. Each SD intervention, coupled with school closures, begins with the start of tighter lockdown, when cumulative incidence exceeds 400 cases (b: inset). The alignment between simulated days and actual dates may slightly differ across separate runs. Case isolation and home quarantine are in place from the outset. } \vspace*{-1mm}
    \label{fig1}
\end{figure}

Using the ABM calibrated to the Delta (B.1.617.2) variant, we varied the macro- and micro-parameters (for CI, HQ, SC and SD), aiming to match the incidence data recorded during the Sydney outbreak in a \emph{retrodiction} mode. As shown in Fig.~\ref{fig1}, the modelling horizon was set to July 25 and assumed a progressive vaccination rollout in addition to a tighter lockdown being imposed at 400 cases (corresponding to July 9).  Construction works were temporarily paused across Greater Sydney during 19--30 July 2021 (inclusive), with the temporary ``construction ban'' lifted on 28 July~\citep{ph-28jul-2021,construction-2021}. Within the considered timeline, the actual incidence growth rate has reduced from $\beta_{I} = 0.098$ (17 June -- 13 July), to $\beta_{II} = 0.076$ (17 June -- 25 July), to $\beta_{III} = 0.037$ (16--25 July), as detailed in Supplementary Material: Growth rates.

The closest match to the actual incidence data over the entire period was produced by a moderate macro-level of social distancing compliance, $SD = 0.5$, or even a lower level ($SD = 0.4$) for the period up to 13 July (see Fig.~\ref{fig1} and Supplementary Material: Sensitivity of outcomes for moderate restrictions, Fig.~S2; also see Section \ref{sec:Disc} for a comparison of these SD levels with real-world mobility reductions).  The match is not exact --- with the actual incidence growth rate changing several times during this period --- perhaps as a consequence of restrictions being imposed heterogeneously across different local government areas. Importantly, however, the growth in actual incidence during the period of the comprehensive lockdown restrictions (16--25 July) is best matched by a higher compliance level, $SD = 0.6$. This match is also reflected by proximity of the corresponding growth rate $\beta_{0.6} = 0.029$ to the incidence growth rate $\beta_{III} = 0.037$.  
The considered SD levels were based on moderately reduced interaction strengths within community, i.e., $SD_c = 0.25$, see Table \ref{tab1}, which were inadequate for  outbreak suppression even with high macro-distancing such as $SD = 0.7$.

\begin{figure}[t] 	
    \centering
    \includegraphics[clip, trim=4.2cm 5.4cm 4.2cm 5.9cm, width=\textwidth]{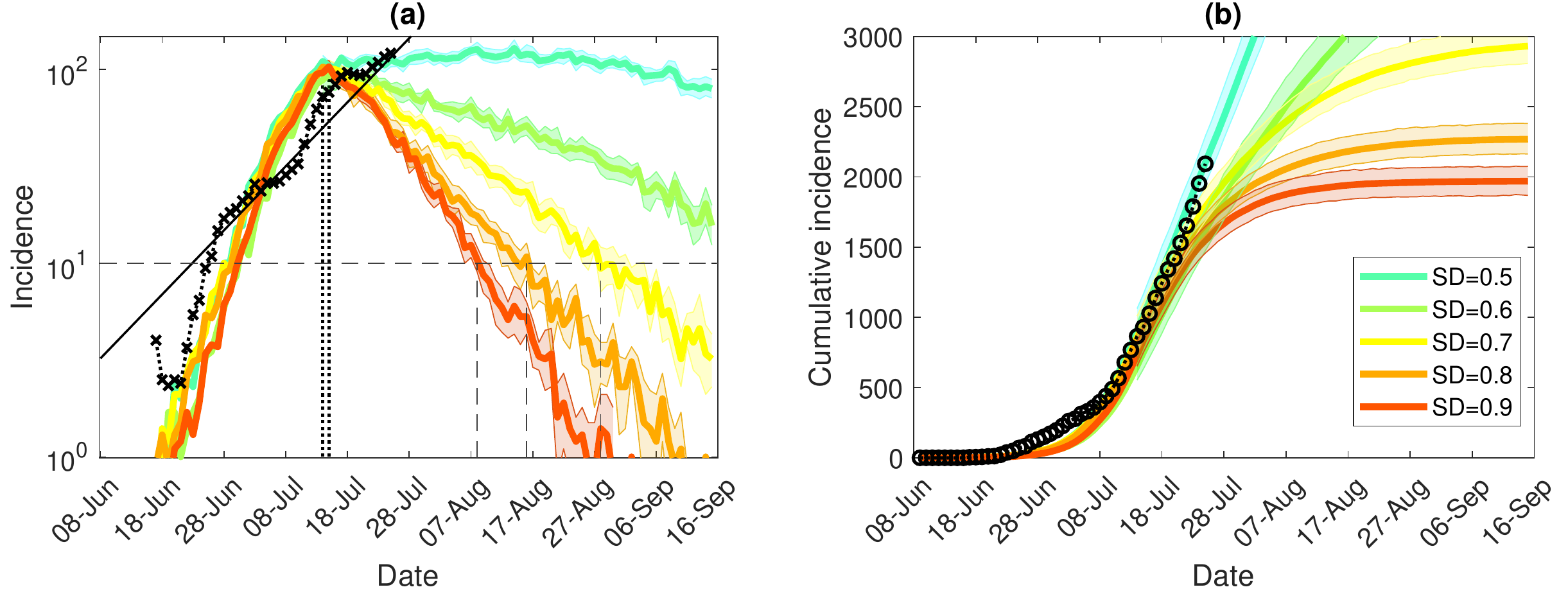}
		
    \caption{\textbf{Tight restrictions (NSW; progressive vaccination rollout; suppression threshold: 400 cases)}: counter-factual simulation scenarios, under lowest feasible interaction strengths ($CI_c = CI_w = 0.1$, $HQ_c = HQ_w = 0.1$, $SD_c = 0.1$, $SC = 0.1$), 
		for (a) (log scale) incidence (crosses), and (b) cumulative incidence (circles). Traces corresponding to feasible social distancing (SD) compliance levels are shown as average over 10 runs (coloured profiles for SD varying in increments of 10\%, i.e., between $SD = 0.5$ and $SD = 0.9$). Vertical lines mark the incidence peaks (dotted) and reductions below 10 daily cases (dashed). Each SD intervention, coupled with school closures, begins with the start of tighter lockdown, when cumulative incidence exceeds 400 cases (i.e., simulated day 31). The alignment between simulated days and actual dates may slightly differ across separate runs. Case isolation and home quarantine are in place from the outset. }
    \label{fig2}
\end{figure}

Furthermore, we considered moderate-to-high macro-levels of social distancing, $0.5 \leq SD \leq 0.9$, while maintaining $CI = 0.7$ and $HQ = 0.5$, in a \emph{counter-factual} mode by reducing the micro-parameters (the interaction strengths for CI, HQ, SC and SD) within their feasible bounds. Again, the control measures were triggered by cumulative incidence exceeding 400 cases (corresponding to a tighter lockdown imposed on July 9). An effective suppression of the outbreak within a reasonable timeframe is demonstrated for macro-distancing at $SD \geq 0.7$, coupled with the lowest feasible interaction strengths for most interventions, i.e., $NPI_c = 0.1$ (where NPI is one of CI, HQ, SC and SD), as shown in Fig.~\ref{fig2} and summarised in Table \ref{tab1}. For $SD = 0.8$, new cases fall below 10 per day approximately a month (33 days) after the peak in incidence, while for $SD = 0.7$ this period reaches 45 days\footnote{A post-peak period duration for each SD level is obtained using the incidence trajectory averaged over ten simulation runs.}. 
Social distancing at $SD = 0.9$ is probably infeasible (as this assumes that 90\% of the population consistently stays at home), but would reduce the new cases to below 10 a day within four weeks (25 days) following the peak in incidence.

\bgroup 
\def\arraystretch{1.3}
\begin{table}[!h]
	\caption{The macro-distancing parameters and interaction strengths: retrodiction (``moderate'') and counter-factual (``tight'').}
	\label{tab1}
	\vspace{5mm}
	\centering
\resizebox{0.8\textwidth}{!}{%
	{\raggedright
	 \noindent
	 \begin{tabular}{l|c|c|c|c}
	 & Macro-distancing & \multicolumn{3}{c}{Interaction strengths}  \\ \hline
	Intervention & Compliance levels & Household & Community & Workplace/School  \\
	 & moderate $\rightarrow$ high &  & moderate $\rightarrow$ tight & moderate $\rightarrow$ tight  \\
	\hline \hline
	CI & 0.7 & 1.0 & 0.25 $\rightarrow$ 0.1 & 0.25 $\rightarrow$ 0.1 \\
  HQ & 0.5 & 2.0 & 0.25 $\rightarrow$ 0.1 & 0.25 $\rightarrow$ 0.1 \\
	SC (children) & 1.0 & 1.0 & 0.5 $\rightarrow$ 0.1 & 0  \\
	SC (parents) & 0.5 & 1.0 & 0.5 $\rightarrow$ 0.1 & 0  \\
	SD & 0.4 $\rightarrow$ 0.8 & 1.0 & 0.25 $\rightarrow$ 0.1 & 0.1  \\
	\hline 
	\end{tabular}
	}
}
\end{table}
\egroup

\bgroup 
\def\arraystretch{1.3}
\begin{table}[!h]
	\caption{Comparison of control measures: projected lockdown duration after the incidence peak, until new cases fall below 10 per day.}
	\label{tab-comp}
	\vspace{5mm}
	\centering
\resizebox{0.8\textwidth}{!}{%
	{\raggedright
	 \noindent
	 \begin{tabular}{c|c|c|c|c|c}
	Vaccination & Vaccination & Lockdown trigger  & \multicolumn{3}{c}{Post-peak duration (days) }  \\ \cline{4-6}
	 scenario &  uptake & (cumulative cases) & $SD = 0.7$ & $SD = 0.8$ & $SD = 0.9$ \\
	\hline \hline
	Pre-pandemic & 6\% & 100 & 55 & 28 & 17 \\
    Progressive & $\rightarrow$ 40\% & 400 & 45 & 33 & 25 \\
	\hline 
	\end{tabular}
	}
}
\end{table}
\egroup

Supplementary Material (Sensitivity of suppression outcomes for tight restrictions) presents results obtained for the scenarios which assume a limited pre-pandemic vaccination coverage (immunising 6\% of the population). A positive impact of the partial progressive rollout which covers up to 40\% of the population by mid-September is counterbalanced by a delayed start of the tighter lockdown, with the 12-day delay leading to a higher peak-incidence, as can be seen by comparing Fig.~\ref{fig2} and Fig.~S4. For example, for $SD = 0.8$, a scenario following the limited pre-pandemic vaccination, but imposing control measures earlier, demonstrates a reduction of incidence below 10 daily cases in four weeks after the peak in incidence (Fig.~S4), rather than 33 days under progressive rollout (Fig.~\ref{fig2}). For $SD = 0.9$ the suppression periods differ by about one week: 17 days (Fig.~S4) against 25 days (Fig.~\ref{fig2}). However, this balance is nonlinear, as shown in Table~\ref{tab-comp}: for $SD = 0.7$, the suppression period under the pre-pandemic vaccination scenario approaches 55 days (Fig.~S4), in contrast to the progressive rollout scenario achieving suppression earlier, in 45 days (Fig.~\ref{fig2}). This is, of course, explained by the longer suppression period under $SD = 0.7$, during which a progressive rollout makes a stronger impact.

\begin{figure}[t]
    \centering
    \includegraphics[clip, trim=2cm 4.cm 1.5cm 4.5cm, width=\textwidth]{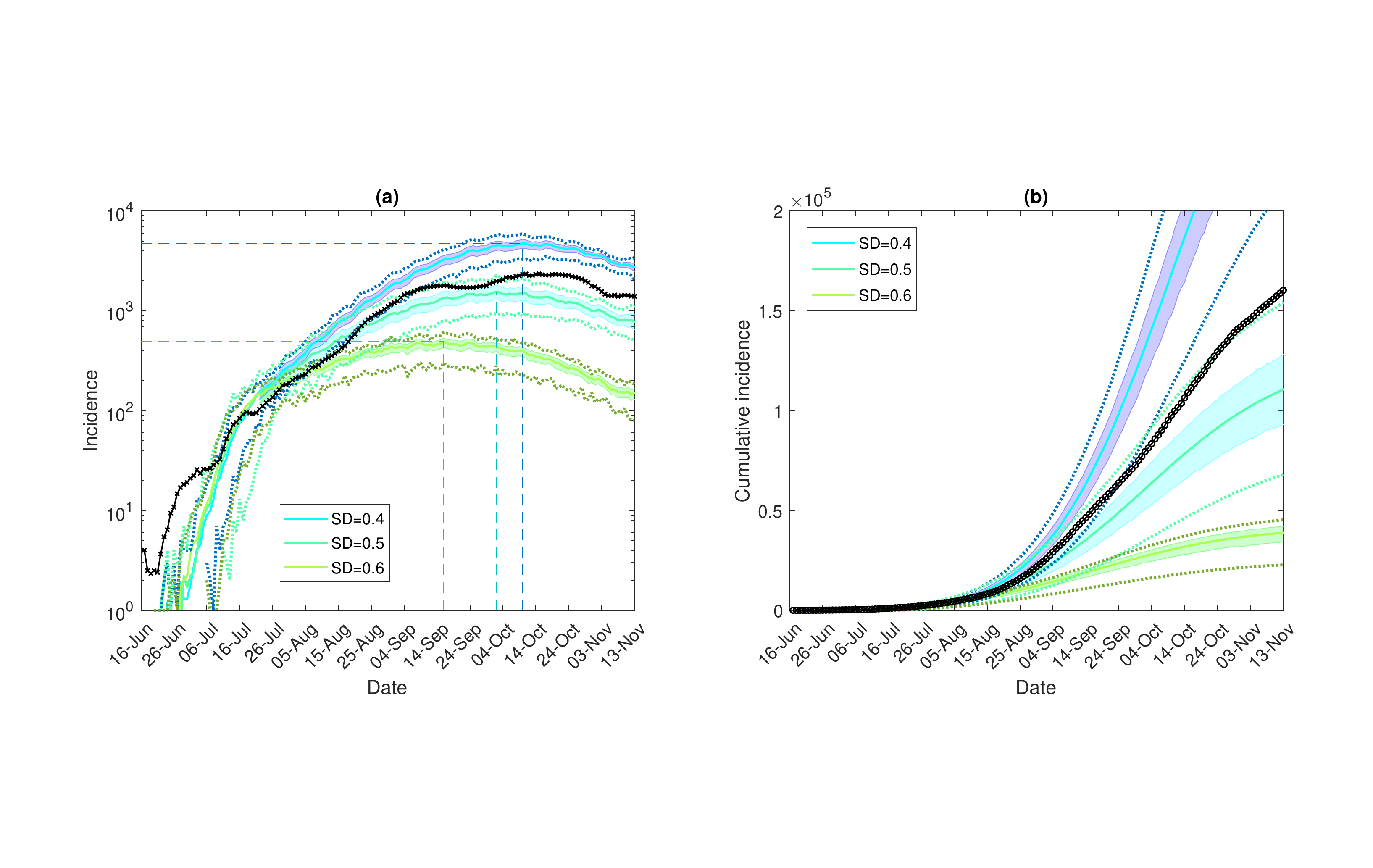} 
    \caption{\textbf{Moderate restrictions (Australia; progressive vaccination rollout; suppression threshold: 400 cases)}: a comparison between simulation scenarios and actual epidemic curves up to November 13, under moderate interaction strengths ($CI_c = CI_w = 0.25$, $HQ_c = HQ_w = 0.25$, $SD_c = 0.25$, $SC = 0.5$). A moving average  of the actual time series for (a) (log scale) incidence (crosses), and (b) cumulative incidence (circles). Traces corresponding to social distancing levels $SD \in \{0.4, 0.5, 0.6\}$  are shown for the period between 16 June and 13 November, as averages over 10 runs (coloured profiles). 95\% confidence intervals are shown as shaded areas. For each SD level, minimal and maximal traces, per time point, are shown with dotted lines. Peaks formed during the suppression period for each SD profile are identified with coloured dashed lines. Each SD intervention, coupled with school closures, begins with the start of initial restrictions. The alignment between simulated days and actual dates may slightly differ across separate runs. Case isolation and home quarantine are in place from the outset.}
    \label{fig-incidence}
\end{figure}

We then considered feasible scenarios tracing the epidemic spread at the national level for the period between mid-June and mid-November 2021, constrained by moderate levels of social distancing, $SD \in \{0.4, 0.5, 0.6\}$, under partial CTNPIs ($CI = 0.7$ and $HQ = 0.5$), see Supplementary Table~S4. A progressive vaccination rollout was simulated concurrently with the continuing restrictions (see Supplementary Material: Vaccination modelling). 
Our Australia-wide model was calibrated by 31 August 2021, adopting a higher fraction of symptomatic children, $\sigma_c = 0.268$ (see Supplementary Material: Model calibration). The actual incidence curve is traced between the profiles formed by $SD = 0.4$ and $SD = 0.5$, with the latter providing the best match. The model projection for incidence peaking across the nation in the range between approximately 1,500 and 5,000 daily cases pointed to early to mid-October. This projection is validated by the actual profiles, as shown in Fig.~\ref{fig-incidence} and Supplementary Material, Fig.~S11. The corresponding levels of simulated and actual vaccination coverage reached across Australia are shown in Supplementary Material: Vaccination modelling.

Using the Australia-wide model, we quantified the expected demand in terms of hospitalisations (occupancy) and the intensive care units (ICUs), and the number of potential fatalities across the nation. The estimation methods are described in Supplementary Material: Hospitalisations and fatalities. The projections obtained for the three feasible levels of social distancing, $SD \in \{0.4, 0.5, 0.6\}$, are shown in Supplementary Figures~S8,~S9, and~S10, and summarised in Table~\ref{tab-hosp} and Supplementary Tables~S9 and S10. The scenario developing under $SD = 0.5$  offers the best match with the actual dynamics again. As expected, the unvaccinated cases form a vast majority among the hospitalisations, ICU occupancy and fatalities (cf. Supplementary Tables~S9 and S10). Importantly, a comparison across the three moderate levels of social distancing, $SD \in \{0.4, 0.5, 0.6\}$ shows that with a 10\% increase in the level of social distancing, the hospitalisations and ICU demand reduce approximately three-fold, and the fatalities reduce at least two times.  These effects of a 10\% increase in the social distancing adherence on the clinical burden and the potential fatalities are robust with respect to changes in the vaccine efficacy against infectiousness, as shown in Supplementary Fig.~S12 and Tables~S9 and S10.

\bgroup 
\def\arraystretch{1.3}
\begin{table}[!t]
	\caption{Estimates (across Australia) of the peak demand in hospitalisations and ICUs; and cumulative fatalities (15 October 2021).}
	\label{tab-hosp}
	\centering
\resizebox{0.7\textwidth}{!}{%
	{\raggedright
	 \noindent
	 \begin{tabular}{l|c|c|c}
	Scenario & Peak hospitalisations: & Peak ICU demand: & Cumulative fatalities: \\  
		 &  mean and 95\% CI &  mean and 95\% CI &  mean and 95\% CI \\ 
	\hline \hline
$SD = 0.4$ & 4805 [4282, 5257] & 812 [731, 885] & 1201 [1057, 1326] \\	
$SD = 0.5$ & 1604 [1358, 1844] & 272 [230, 312] & 539 [479, 624] \\	
$SD = 0.6$ & 533 [476, 579] & 91 [80, 99] & 235 [209, 256]  \\
	\hline
  Actual & 1551 (28 September) & 308 (12 October) & 596 (15 October)   \\	
	\hline 
	\end{tabular}
	}  
}
\end{table} 
\egroup

\vspace{-2mm}
\section*{Discussion}
\label{sec:Disc}

Despite a relatively high computational cost, and the need to calibrate numerous internal parameters, ABMs capture the natural history of infectious diseases in a good agreement with the established estimates of incubation periods, serial/generation intervals, and other key epidemiological variables. Various ABMs have been successfully used for simulating actual and counter-factual epidemic scenarios based on different initial conditions and intervention policies~\citep{germann2006mitigation,ajelli2010comparing,nsoesie2012sensitivity,zachreson2020interfering,milne2021reliance}. 

Our early COVID-19 study~\citep{chang2020modelling} modelled transmission of the ancestral lineage of SARS-CoV-2 characterised by the basic reproduction number of $R_0 \approx 3.0$ (adjusted $R_0 \approx 2.75$).  
This study compared several NPIs and identified the minimal SD levels required to control the first wave in Australia. Specifically, a compliance at the 90\% level, i.e., $SD = 0.9$ (with $SD_w = 0$ and $SD_c = 0.5$) was shown to control the disease within 13-14 weeks. This relatively high SD compliance was required in addition to other restrictions (TR, CI, HQ), set at moderate levels of both macro-distancing ($CI = 0.7$ and $HQ = 0.5$), and interaction strengths: $CI_w = HQ_w = CI_c = HQ_c = 0.25$, $CI_h = 1.0$, and $HQ_h = 2.0$~\citep{chang2020modelling}.

The follow-up work~\citep{zachreson2021how} quantified possible effects of a mass-vaccination campaign in Australia, by varying the extents of \emph{pre-pandemic} vaccination coverage with different vaccine efficacy combinations. This analysis considered hybrid vaccination scenarios using two vaccines adopted in Australia: BNT162b2 (Pfizer/BioNTech) and ChAdOx1 nCoV-19 (Oxford/AstraZeneca). Herd immunity was shown to be out of reach even when a large proportion (82\%) of the Australian population is vaccinated under the hybrid approach, necessitating future partial NPIs for up to 40\% of the population. The model was also calibrated to the basic reproduction number of the ancestral lineage ($R_0 \approx 3.0$, adjusted $R_0 \approx 2.75$), and used the same moderate interaction strengths as the initial study~\citep{chang2020modelling} (except $SD_c = 0.25$, reduced to match the second wave in Melbourne in 2020).  

In this work, we re-calibrated the ABM to incidence data from the ongoing third pandemic wave in Australia driven by the Delta variant. The reproductive number was calibrated to be at least twice as high ($R_0 = 5.97$) as the one previously estimated for pandemic waves in Australia. We then explored effects of available NPIs on the outbreak suppression, under a \emph{progressive vaccination} scenario. The retrodictive modelling identified that the current epidemic curves, which continued to grow (until mid-October 2021), can be closely matched by moderate social distancing coupled with moderate interaction strengths within community ($SD$ in $[0.4, 0.5]$, $SD_c = 0.25$), as well as moderate compliance with case isolation ($CI = 0.7$, $CI_w = CI_c = 0.25$) and home quarantine ($HQ = 0.5$, $HQ_w = HQ_c = 0.25$). The estimate of compliance has briefly improved to $SD \approx 0.6$ during the period of comprehensive lockdown measures, announced on July 17, but returned to $SD \approx 0.5$ in early August. 

We note that the workers delivering essential services are exempt from lockdown restrictions. The fraction of the exempt population can be inferred conservatively as 4\% (strictly essential)~\citep{wang2020global}, more comprehensively as approximately 19\% (including health care and social assistance; public administration and safety; accommodation and food services; transport, postal and warehousing; electricity, gas, water and waste services; financial and insurance services), but can reach more significant levels, around 33\%, if all construction, manufacturing, and trade (retail/wholesale) are included in addition~\citep{industry2019}. The latter, broad-range, case limits feasible social distancing levels to approximately $SD \approx 0.7$.  However, even with these inclusions, there is a discrepancy between the level estimated by ABM ($SD$ in $[0.4, 0.5]$) and the broad-range feasible level ($SD \approx 0.7$). This discrepancy would imply that approximately 20-25\% of the population have not been consistently complying with the imposed restrictions, while 30-35\% may have been engaged in services deemed broadly essential (other splits comprising 50-60\% of the ``non-distancing'' population are possible as well). 

The inferred levels of social distancing are supported by real-world mobility data~\citep{mobility-jul-2021}. Specifically, when compared to baseline (i.e., the median value for the corresponding day of the week, during the five-week period 3 January -- 6 February 2020, as set by data provider to represent the pre-pandemic levels), the reports for July 16 showed 31\% reduction of mobility at workplaces, and 37\% reduction of mobility in retail and recreation settings, with concurrent 65\% reduction of mobility on public transport. On July 21, the mobility reductions were reported as 43\% (workplaces), 41\% (retail and recreation), and 72\% (public transport). The extent of the mobility reduction in workplaces, as well as retail and recreation, closely matched the social distancing levels estimated by the model (approximately 40\%). The partial reductions in mobility across workplaces, retail and recreation have since been maintained around 40-50\% on average~\citep{mobility-jul-2021}. According to numerous reports~\citep{essential-nsw-2021,essential-vic-2021,nsw-oct-2021}, the infection spread among essential workers was substantial, and the interactions within workplaces and community contributed to the disease transmission stronger than contacts in public transport. 

Moderate levels of compliance ($SD$ in $[0.4, 0.6]$) would be inadequate for suppression of even less transmissible coronavirus variants~\citep{chang2020modelling}. The Delta variant demands a stronger compliance and a reduction in the scope of essential services (especially, in a setting with low immunity).  Specifically, our results indicate that an effective suppression within a reasonable timeframe can be demonstrated only for very high compliance with social distancing ($SD \geq 0.7$), supported by dramatically reduced, and practically infeasible, interaction strengths within the community and work/study environments ($NPI_c = NPI_w = 0.1$).  Importantly, a significant fraction of local transmissions during the Sydney outbreak in NSW, as well as during the following outbreak in Melbourne, VIC (which started on 13 July 2021, was initially suppressed, but then resumed its growth on 4 August 2021~\citep{covid19data}), occurred in the suburbs characterised by socioeconomic disadvantage profiles, as defined by The Australian Bureau of Statistics’ Index of Relative Socio-economic Advantage and Disadvantage (IRSAD)~\citep{essential-nsw-2021,essential-vic-2021,nicholas2021most}. To a large extent, the epidemic spread in these suburbs was driven by structural factors, such as higher concentrations of essential workers, high-density housing, shared and multi-generational households, etc. Thus, even a combination of government actions (e.g., a temporary inclusion of some services previously deemed essential under the lockdown restrictions~\citep{ph-28jul-2021,construction-2021}, while providing appropriate financial support to the affected businesses and employees), and a moderate community engagement with the suppression effort, proved to be insufficient for the outbreaks' suppression.  

Obviously, the challenges of suppressing emerging variants of concern can be alleviated by a growing vaccination uptake. However, in Australia, the vaccination rollout was initially limited  by various supply and logistics constraints. Furthermore, as our results demonstrate, a progressive vaccination rollout reaching up to 40\% of the population (i.e., approximately 50\% of adults) was counter-balanced by a delayed introduction of the tighter control measures. This balance indicated that a comprehensive mass-vaccination rollout plays a crucial role over a longer term and should preferably be carried out in a pre-outbreak phase~\citep{zachreson2021how}. Ultimately, the epidemic peak in NSW during the lockdown period was reached  only when about a half of the adults were double vaccinated by mid-September (i.e., 49.6\% on 15 September 2021)~\citep{aus--2021}. Across the nation, the peak in incidence was observed by mid-October (as predicted by the model), once approximately two thirds of adults were double vaccinated~\citep{aus--2021}, also in concordance with the model (see Supplementary Material: Vaccination modelling). 

A post-lockdown increase in infections is expected when the stay-at-home orders are lifted in recognition of immunising 70\%, and then 80\%, of adults~\citep{burnet-18-sep-2021}. However, a detailed analysis of a possible post-lockdown surge in infections, the resultant increased demand on the healthcare system, and potential fatalities, is outside of the scope for this study.

While the model was not directly used to inform policy, it forms part of the information set available to health departments, and we hope that its policy relevance can contribute to rapid and comprehensive responses in jurisdictions within Australia and overseas. 
A failure in reducing the size of the initial outbreak, due to a delayed vaccination rollout, challenging socioeconomic profiles of the primarily affected areas, inadequate population compliance, and a desire to maintain and restart socioeconomic activities, has generated a substantial pandemic wave affecting the entire nation~\citep{viana2021controlling,burnet2021,ouakrim2021}. 

\subsection*{Study limitations}

In modelling the progressive vaccination rollout, we assumed a constant weekly uptake rate of 3\%, while the rollout was accelerating. The rate of progressive vaccination is expected to vary, being influenced by numerous factors, such as access to national stockpiles, dynamics of social behaviour, and changing medical advice.  In addition, we did not consider a diminishing vaccine efficacy, given that the temporal scope of the study was limited to a relatively short period of 6 months (June--November 2021) during which a progressive rollout was modelled. Thus, only a relatively small fraction of the population vaccinated during the very first few months would be experiencing a tangibly diminished vaccine efficacy (with respect to the Delta variant)~\cite{pouwels2021effect}. Nevertheless, the study included a sensitivity analysis of the vaccine efficacy across three static levels.

Another limitation is that the surrogate ABM population which corresponds to the latest available Australian Census data from 2016 (23.4M individuals, with 4.45M in Sydney) is smaller than the current Australian population (25.8M, with 4.99M in Sydney). We expect low sensitivity of our results to this discrepancy due to the outbreak size being three orders of magnitude smaller than Sydney population. 

Finally, the model does not directly represent in-hotel quarantine and in-hospital transmissions. Since the frontline professionals (health care and quarantine workers) were vaccinated in a priority phase carried out in Australia in early 2021, i.e., before the Sydney outbreak, this limitation is expected to have a minor effect. 
Overall, as the epidemiology of the Delta variant continues to be refined with more data becoming available, our results may benefit from a retrospective analysis.

%

\section*{Author Contributions}

MP conceived and co-supervised the study and drafted the original Article. SLC and MP designed the computational experiments,  re-calibrated the model, and estimated hospitalisations, ICU occupancy, and potential fatalities. CZ implemented simulations of progressive vaccination and social distancing policies. SLC carried out the computational experiments, verified the underlying data, and prepared all figures. All authors had full access to all the data in the study.  All authors contributed to the editing of the Article, and read and approved the final Article.

\section*{Funding}
This work was partially supported by the Australian Research Council grants DP220101688 and DP200103005 (MP and SLC). 
Additionally, CZ is supported in part by National Health and Medical Research Council project grant (APP1165876). 
AMTraC-19 is registered under The University of Sydney’s invention disclosure CDIP Ref. 2020-018. 

\section*{Acknowledgments}
We are thankful for support provided by High-Performance Computing (HPC) service (Artemis) at the University of Sydney.

\section*{Data Availability Statement}
We used anonymised data from the 2016 Australian Census obtained from the Australian Bureau of Statistics (ABS). These datasets can be obtained publicly, with the exception of the work travel data which can be obtained from the ABS on request. It should be noted that some of the data needs to be processed using the TableBuilder: https://www.abs.gov.au/websitedbs/censushome.nsf/home/tablebuilder.  The actual incidence data are available from the health departments across Australia (state, territories, and national), and at: https://www.covid19data.com.au/. Other source and supplementary data, including simulation output files, are available at Zenodo~\citep{amtrac-data-zenodo-2021}. The source code of AMTraC-19 is also available at Zenodo~\citep{amtrac-code-zenodo-2021}.


\newpage

\newcommand{\beginsupplement}{%

 \setcounter{table}{0}
   \renewcommand{\thetable}{S\arabic{table}}%
   
     \setcounter{figure}{0}
      \renewcommand{\thefigure}{S\arabic{figure}}%
      
      \setcounter{page}{1}
      \renewcommand{\thepage}{S\arabic{page}} 
      
      \setcounter{section}{0}
      \renewcommand{\thesection}{S\arabic{section}}
      
      \setcounter{equation}{0}
      \renewcommand{\theequation}{S\arabic{equation}}
     }

\doublespace

\beginsupplement

\section*{Supplementary Material}

\section*{Agent-based Model of Transmission and Control of the COVID-19 pandemic in Australia: AMTraC-19}

\subsection*{Demographics}

Each agent in the artificial population belongs to several mixing groups stochastically generated from census data based on Statistical Areas (SA1 and SA2) level statistics, and the distributions across age groups, households and workplaces~\citep{cliff2018nvestigating,zachreson2018urbanization,fair2019creating,chang2020modelling}.  Agents are split into five different age groups: preschool aged children (0-4), children (5-18), young adults (19-29), adults (30-64) and older adults (65+), with a further refinement into specific ages derived when necessary from the census distribution.  During the daytime simulation cycle (time-step), agents interact in ``work regions'', e.g., an agent representing an adult individual (19-64) interacts within a work group, while children agents (5-18) interact within classrooms, grades, and schools. During the nighttime cycle, individuals interact in ``home regions'', e.g., households, household clusters, neighbourhoods (SA1), and communities (SA2). Preschool children and older adults interact only in home regions during the nighttime simulation cycle. During weekends, the nighttime simulation cycle runs twice, thus replacing daytime interactions in work regions with an additional interaction cycle in home regions.

\subsection*{Transmission probability}

At each time-step $n$ the simulator computes the probability of infection $p_i(n)$ for a susceptible agent $i$. This is determined by considering all relevant mixing contexts (daytime or nighttime) $g$ for the agent $i$, selected from $G_i(n)$, and the infection states of other agents $j$ in each context $A_g$. The context-dependent probability that infectious individual $j$ infects susceptible individual $i$ in context $g$ in a single time step, $p^g_{j \rightarrow i}$, is defined as follows:
\begin{equation} \label{eq:prob-transmission}
	p_{j \to i}^g(n ) = \kappa \ f( n - n_j \mid j ) \ q_{j \to i}^g
\end{equation}
where $\kappa$ is a global scaling factor (selected to calibrate to the reproductive number $R_0$), $n_j$ denotes the time when agent $j$ becomes infected, and $q_{j \to i}^g$ is the probability of transmission from agent $j$ to $i$ at the infectivity peak, derived from the  transmission or contact rates. The function $f : \mathbb{N} \to [0,1]$ quantifies the infectivity of agent $j$ over time, according the natural history of the disease; $f( n - n_j \mid j ) = 0$ when $n < n_j$; cf. Supplementary Fig.~\ref{fig3}.
The transmission rates $q_{j\to i}^g$ for the household and study environments are shown in Table~\ref{tab:transmission-rates}. The contact rates $c_{j \to i}^g$ for household clusters, neighbourhoods, and communities are detailed in Table~\ref{tab:contact-table}. These contact rates are rescaled, using a fixed scaling factor $\rho$, to transmission rates~\citep{cliff2018nvestigating}:
\begin{equation} \label{eq:ptrans}
	q_{j \to i}^g = \rho \ c^g_{j \to i}.
\end{equation}
The overall probability that a susceptible agent $i$ is infected at a given time step $n$ is then calculated as~\citep{chang2020modelling}
\begin{equation}
    p_i(n) = 1 - \prod_{g \in G_i(n)} \left[  \prod_{j \in A_g\setminus i} (1 - p^g_{j \rightarrow i}(n))  \right].
		\label{eq}
\end{equation}
This expression is adjusted to account for the agents adhering to various non-pharmaceutical interventions and vaccinations, as detailed below, see (\ref{eq-fg}) and (\ref{eq-ve}). At the end of each cycle, a Bernoulli trial with probability $p_i(n)$ determines if a susceptible agent becomes infected.

\subsection*{Natural history of disease}

In a single agent, the disease progression from exposure to recovery develops over several agent states: \textsc{susceptible}, \textsc{latent}, infectious \textsc{symptomatic}, infectious \textsc{asymptomatic}, and \textsc{recovered}. In general, the first phase is the latent period during which infected agents are unable to infect others. However, in modelling the Delta variant, AMTraC-19 sets this period to zero days. The second phase is the incubation period characterised by an exponentially increasing infectivity, from 0\% to 100\%, reaching its peak at the end of the incubation period after $T_{inc}$ days (see Supplementary Fig.~\ref{fig3}).
 In the third, post-incubation, phase the infectivity decreases linearly from the peak to zero, over $T_{rec}$ days until the recovery (with immunity). The parameters $T_{inc}(i)$ and $T_{rec}(i)$ are randomly generated for each agent $i$, see Table~\ref{tab-param}, thus defining the disease progression in the affected agent, i.e., $D(i) = T_{inc}(i) + T_{rec}(i)$.

Asymptomatic cases are set to be 50\% as infectious as symptomatic cases, $\alpha = 0.5$. We assume that 67\% of adult cases are symptomatic ($\sigma_a = 0.67$), and a lower fraction (either 13.4\% or 26.8\%) is set as symptomatic in children (e.g., $\sigma_c = 0.268$). The fractions $\sigma_{a,c}$ reduce the probability of becoming ill (symptomatic) $p^d_i(n)$, given the infection probability $p_i(n)$, for each adult or child agent: $p^d_i(n) = \sigma_{a|c} p_i(n)$.
On each simulated day, pre-symptomatic, symptomatic and asymptomatic cases are detected with specific probabilities $r$ (see Supplementary Material: Model calibration). Only detected cases are counted in the incidence profiles. Table~\ref{tab-param} summarises the main model parameters.

\subsection*{Reproductive number}

For every scalar $\kappa$, the reproductive number $R_0$ is estimated numerically~\citep{zachreson2020interfering}, by stochastic sampling of index cases (sample size $\approx 10^4$). Every micro-simulation randomly selects a single index case, and detects the number of secondary infections generated during the period until the index case is recovered. The secondary cases themselves are prevented from generating further infections, so that all detected cases are attributed to the index case. In order to reduce the bias in selecting a typical (rather than purely random) index case, we employ ``the attack rate pattern weighted index case'' method~\citep{germann2006mitigation,zachreson2020interfering}, based on age-specific attack rates. These age-stratified weights, computed as averages over many full simulation runs, are assigned to secondary cases produced by the micro-simulation sample of index cases. This accounts for the correlations between age groups and population structure~\citep{miller2009spread}.  Given the five age groups, [0--4, 5--18, 19--29, 30--64, 65+], the following age-specific weights were used in producing $R_0$ as the weighted average of secondary cases: [0.064, 0.1919, 0.1412, 0.4583, 0.1446].  

This procedure is different from an empirical estimation of the reproductive number $R_0$ which would require comprehensive and still unavailable data on the actual secondary infections generated by different index cases. Instead, the numerical estimations of $R_0$ derived from the ABM are compared with the known estimates of $R_0$, and further validated by checking the concordance between projected and actual epidemic curves (see Supplementary Material: Model calibration).

\subsection*{Non-pharmaceutical interventions}

The agents affected by various NPIs (case isolation: CI; home quarantine: HQ; school closures: SC; social distancing: SD) are determined in the beginning of each simulation run, given specific compliance levels explored by a simulation scenario.  
Every intervention F is specified via the fraction $F$ of the population complying with the NPI (``macro-distancing''), and a set of interaction strengths $F_g$ (``micro-distancing'') that modify the transmission probabilities within a specific mixing context $g$: households ($F_h$), communities ($F_c$), and workplaces/study environments ($F_w$). 
For non-complying agents $j$, the interaction strengths are unchanged, i.e., $F_g(j) = 1$, while for complying agents $j$, the strengths are generally different: $F_g(j) \neq 1$, so that the transmission probability of infecting a susceptible agent $i$ is adjusted as follows:
\begin{equation}
    p_i(n) = 1 - \prod_{g \in G_i(n)} \left[  \prod_{j \in A_g\setminus i} (1 - F_g(j) \ p^g_{j \rightarrow i}(n))  \right].
		\label{eq-fg}
\end{equation}
The intervention-induced restrictions are applied in a specific order: CI, HQ, SD, SC (for parents SC$^a$ and children SC$^c$), with only the most relevant interaction strength $F_g$ applied during each simulation cycle. 
For example, if a symptomatic student is in case isolation, then the interaction strengths $HQ_g$, $SD_g$ and $SC^c_g$ would not modify the agent's transmission probabilities, even if this agent is considered compliant with the corresponding measures, and the only applicable strength would be $CI_g$.
The macro-distancing levels of compliance and the interaction strengths (micro-parameters) defining the NPIs are summarised in Table~1. To re-iterate, the macro-distancing compliance levels across interventions F define how many agents (i.e., $F$) adjust their interaction strengths to the micro-distancing levels $F_g$ within specific contexts $g$.  

At macro-level, some interventions, e.g., the CI and HQ measures, are set to last during the full course of the simulated scenario. The duration of SD and/or SC measures varies. In general, there may be a predefined set of intervals describing the (possibly interrupted and resumed) duration of intervention F. In AMTraC-19, we express the continuous duration as the number of days, $F_T$, following a threshold $F_X$ in cumulative detected cases. At micro-level, the interaction strengths are reduced during the same period $F_t$ for most of the measures, except HQ which is modelled to reduce the interaction strengths of the compliant agents for 14 days. The micro-duration of CI is limited by the disease progression in the affected agent $i$, i.e., $D(i)$.
Formally, an intervention F is defined by a set of parameters: $\{F, F_T, F_X, F_h, F_c, F_w, F_t\}$, for example, SD may be defined by $\{0.6, 191, 400, 1.0, 0.25, 0.1, 191\}$. A scenario is then defined by a combination of these sets defined for all interventions CI, HQ, SD, SC$^a$ and SC$^c$, see Table~\ref{tabX}.

\subsection*{Vaccination modelling}

The national COVID-19 vaccine rollout strategy pursued by the Australian Government follows a hybrid approach combining two vaccines: BNT162b2 (Pfizer/BioNTech) and  ChAdOx1 nCoV-19 (Oxford/AstraZeneca), administered across specific age groups 
(the eligibility policy has underwent multiple changes, with different age groups provided access progressively).
Our model accounts for differences in vaccine efficacy for the two vaccines approved for distribution in Australia, and distinguishes between separate vaccine components: efficacy against susceptibility (\VEs), disease (\VEd) and infectiousness (\VEi). 

Given these components, the transmission probability of infecting a susceptible agent $i$ is adapted as follows:
\begin{equation}
    p_i(n) = (1 - \VEs_i) \left ( 1 - \prod_{g \in G_i(n)} \left[  \prod_{j \in A_g\setminus i} (1 - (1 - \VEi_j) F_g(j) \ p^g_{j \rightarrow i}(n))  \right] \right )
		\label{eq-ve}
\end{equation}
where for vaccinated agents $\VEi_j = \VEi$ and $\VEs_i = \VEs$, and for unvaccinated agents $\VEi_j = \VEs_i = 0$. The probability of becoming ill (symptomatic) is affected by the efficacy against disease (\VEd) as follows: $p^d_i(n) = (1 - \VEd) \sigma_{a|c} p_i(n)$ for adults and children.

For the pre-pandemic vaccination rollout, the extent of pre-outbreak vaccination coverage was set at 6\% of the population, approximately matching the level actually achieved in Australia by mid-June 2021. For the progressive vaccination rollout, the initial coverage was set at zero, followed by vaccination uptake averaging 3\% per week for the duration of simulation, and reaching the levels detailed in Table~\ref{tab-vacc}. Specifically, a policy-relevant milestone of 70\% vaccinated adults is reached within the model around 13 November, that is, after 121 days of social distancing and school closures, see Supplementary Data 2; cf. Table~\ref{tabX} for duration of measures.

In setting the efficacy of vaccines against B.1.617.2 (Delta) variant, we followed the study of Bernal et al.~\citep{bernal2021effectiveness} which estimated the efficacy of BNT162b2 (Pfizer/BioNTech) as VEc $\approx$ 0.9 (more precisely, 87.9\% with 95\% CI: 78.2 to 93.2), and the efficacy of ChAdOx1 nCoV-19 (Oxford/AstraZeneca) as VEc $\approx$ 0.6 (i.e., 59.8\% with 95\% CI: 28.9 to 77.3). 
Given the constraint for the clinical efficacy~\citep{zachreson2021how}:
\begin{equation}\label{Eq_VEc_supp}
    \VEc = \VEd + \VEs  - \VEs \ \VEd,
\end{equation}
we set \VEd = \VEs = 0.684 for BNT162b2, and \VEd = \VEs = 0.368 for ChAdOx1 nCoV-19. 

Recent studies also provided the estimates of efficacy against infectiousness (VEi) for both considered vaccines at a level around 0.5~\citep{harris2021impact}. A general sensitivity analysis of the model to changes in VEi and VEc was carried out in~\citep{zachreson2021how}. 

In both rollout scenarios, the vaccinations are assumed to be equally balanced between the two vaccines, so that each type is given to approximately (i) 0.7M individuals initially, by mid-June, or (ii) 4.7M individuals progressively, by mid-September. Vaccines are distributed according to specific age-dependent allocation ratios, $\approx$ 2547:30,000:1000, mapped to age groups [age $\geq 65$] : [$18 \leq \text{age} < 65$] : [age $< 18$ ], as explained in our prior work~\citep{zachreson2021how}.
That is, for every 2547 agents aged over 64 years, 30,000 individuals aged between 18 and 64 years, and 1000 agents under the age of 18 years are immunised. The allocation ratios are aligned with the age distribution of the Australian population (based on the 2016 ABS Census), while reflecting the tighter regulations on vaccine approval for children. At each simulation cycle this process immunises agents at a fixed rate. The allocations continue over a number of cycles until all adult agents are immunised. At the end of all allocations, the fraction of immunised children reaches approximately 20\% of all children (i.e., agents under the age of 18 years).

\section*{Model calibration}

In order to model transmission of the Delta (B.1.617.2) variant during the Sydney outbreak of COVID-19 (June--July 2021), we re-calibrated the model to match the reproduction number approximately twice as high as our previous estimates ($R_0 \approx 3.0$) for the two waves in Australia in 2020. In aiming at this level, we followed global estimates, which showed that the $R_0$ for B.1.617.2 is increased by 97\% (95\% CI of 76--117\%) relative to its ancestral lineage~\citep{campbell2021increased}. As implemented in our model, the re-calibrated reproductive number was estimated to be $R_0 = 5.97$ with a 95\% CI of 5.93--6.00. The corresponding generation period is estimated to be $T_{gen} = 6.88$ days with a 95\% CI of 6.81--6.94 days. The fraction of symptomatic children among all pediatric cases was set to $\sigma_c = 0.134$. The 95\% confidence intervals (CIs) were constructed from the bias corrected bootstrap distribution~\citep{tibshirani1993introduction}. 

The model calibration varied the scaling factor $\kappa$ (which scales age-dependent contact and transmission rates) in increments of 0.1. The best matching $\kappa$ was identified when the resultant reproductive number, estimated in this work using age-stratified weights~\citep{zachreson2020interfering}, was close to $R_0 = 6.0$.  The procedure resulted in the following parametrisation: 
\begin{itemize}
\item the scaling factor $\kappa = 5.3$ produced  $R_0 = 5.97$ with 95\% CI of 5.93--6.00 ($N = 6318$, randomly re-sampled in 100 groups of 100 samples; confidence intervals constructed by bootstrapping with the bias-corrected percentile method~\citep{tibshirani1993introduction});
\item the fraction of symptomatic cases was set as $\sigma_a = 0.67$ for adults, and $1/5$ of that, i.e., $\sigma_c = 0.134$, for children; 
\item different transmission probabilities for asymptomatic/presymptomatic and symptomatic agents were set as ``asymptomatic infectivity" (factor of 0.5) and ``pre-symptomatic infectivity'' (factor of 1.0)~\citep{ferretti2020quantifying,wu2020estimating}; 
\item incubation period $T_{inc}$ was chosen to follow log-normally distributed incubation times with mean 4.4 days ($\mu = 1.396$ and $\sigma = 0.413$)~\citep{zhang2021transmission}; 
\item a post-incubation infectious asymptomatic or symptomatic period was set to last between 7 and 14 days (uniformly distributed)~\citep{cdc2021interim,arons2020presymptomatic,wolfel2020virological}; and 
\item different detection probabilities were set as symptomatic (detection per day is 0.227) and asymptomatic/pre-symptomatic rates (detection per day is 0.01)~\citep{zachreson2021how}. 
\end{itemize}
Calibration of the fraction of symptomatic cases in children, including its higher setting $\sigma_c = 0.268$, is detailed in Supplementary Material: Sensitivity analysis. Estimation of the growth rates in incidence is described in Supplementary Material: Growth rates. In summary, two parameters were varied during the calibration: the scaling factor $\kappa$ and the fraction of symptomatic cases in children $\sigma_c$, both reflecting specifics of the Delta variant. Other key epidemiological parameters listed above, e.g., different transmission probabilities, were calibrated previously~\citep{chang2020modelling,zachreson2021how}. The remaining parameters describing the natural history of the disease, e.g., the incubation period, were set according to the available epidemiological evidence and varied during a sensitivity analysis, as described below. Table~\ref{tab-calibr} summarises the key calibration outcomes. 

The parametrisation of NPIs is independent of the disease model implemented in AMTraC-19, and thus, the NPI parameters were kept constant during calibration, being varied only within different scenarios (i.e., in setting moderate versus tight restrictions). As a result, the reported findings quantifying specific effects of interventions are decoupled from the disease model calibration.

\section*{Sensitivity analysis}

Several internal parameters have been varied during prior sensitivity analyses~\citep{chang2020modelling,zachreson2021how}. For this study, we carried out additional sensitivity analyses in terms of the incubation period, the reproductive number, the generation period, and the fraction of symptomatic cases for children $\sigma_c$. The analysis presented below covers the time period between 17 June and 13 July inclusively, and is based on the pre-pandemic vaccination rollout. It can be contrasted with the progressive vaccination rollout studied in the main manuscript.

\subsection*{Incubation period} 

While previously the incubation period of COVID-19 was estimated to be distributed with the mean 5.5 days~\citep{ferretti2020quantifying,lauer2020incubation}, a more recent study of the Delta variant reported a shorter mean incubation period: 4.4 days (with 95\% CI of 3.9-5.0)~\citep{zhang2021transmission}. Our previous sensitivity analysis~\citep{chang2020modelling} showed that the model is robust to changes in the time to peak infectivity, investigated in the range between 4 and 7 days. Here we investigated the sensitivity of the updated model to changes in the incubation period specifically, varying it between the mean 4.4 days (log-normally distributed with $\mu = 1.396$ and $\sigma = 0.413$), matching the estimates of Zhang et al.~\citep{zhang2021transmission} and the mean 5.5 days (log-normally distributed with $\mu = 1.644$, $\sigma = 0.363$)~\citep{ferretti2020quantifying}.

\begin{figure}
    \centering
    \includegraphics[clip, trim=1cm 0cm 1cm 0cm, width=\textwidth]{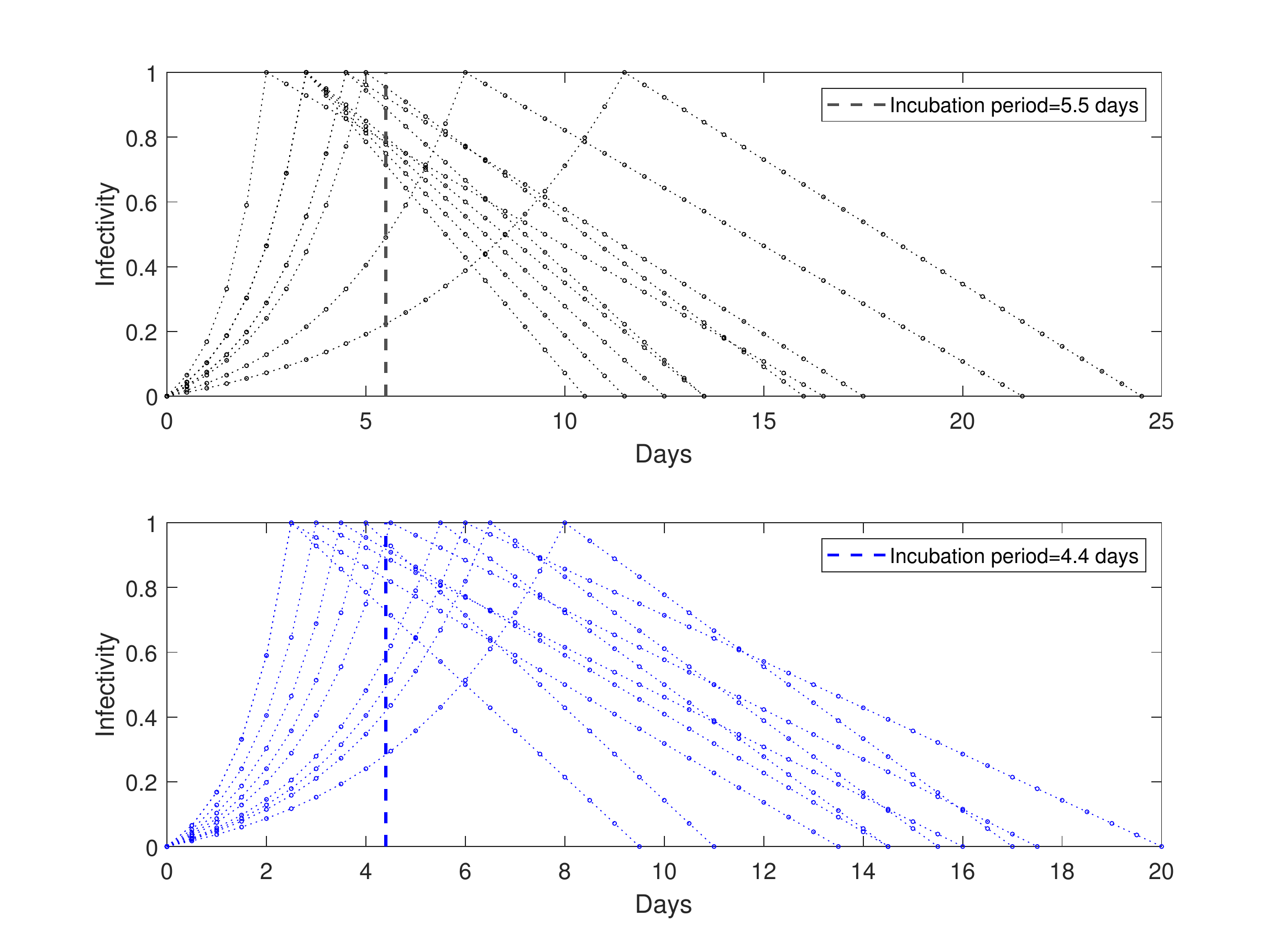}
    \caption{Model of the natural history of COVID-19. Profiles of the infectivity are sampled from 20 random agents, with each profile rising exponentially until a peak, followed by a linear decrease to full recovery. Vertical dashed lines indicate the mean incubation period $T_{inc}$: 5.5 days (top) and 4.4 days (bottom), with the means distributed log-normally. }
    \label{fig3}
\end{figure}

The comparison between the 4.4-day and 5.5-day incubation periods was carried out for the same scaling factor $\kappa = 5.3$. The corresponding reproductive number changed from $R_0 = 5.97$ (95\% CI of 5.93--6.00, $N = 6318$, $T_{inc} = 4.4$) to $R_0 = 6.39$ (95\% CI of 6.36--6.43, $N = 7804$, $T_{inc} = 5.5$), that is, by approximately 7\%. Similarly, the corresponding generation periods changed from $T_{gen} = 6.88$ (95\% CI of 6.81--6.94, $N = 6318$, $T_{inc} = 4.4$) to $T_{gen} = 7.77$ (95\% CI of 7.71--7.83, $N = 7804$, $T_{inc} = 5.5$), i.e., by approximately 13\%. This relatively small sensitivity is explained by the high level of infectivity exhibited in our model by pre-symptomatic and asymptomatic individuals, see. Fig.~\ref{fig3}.

\subsection*{Sensitivity of outcomes for moderate restrictions} 

\begin{figure}
    \centering
    \includegraphics[clip, trim=4.2cm 4.6cm 4.2cm 5cm, width=\textwidth]{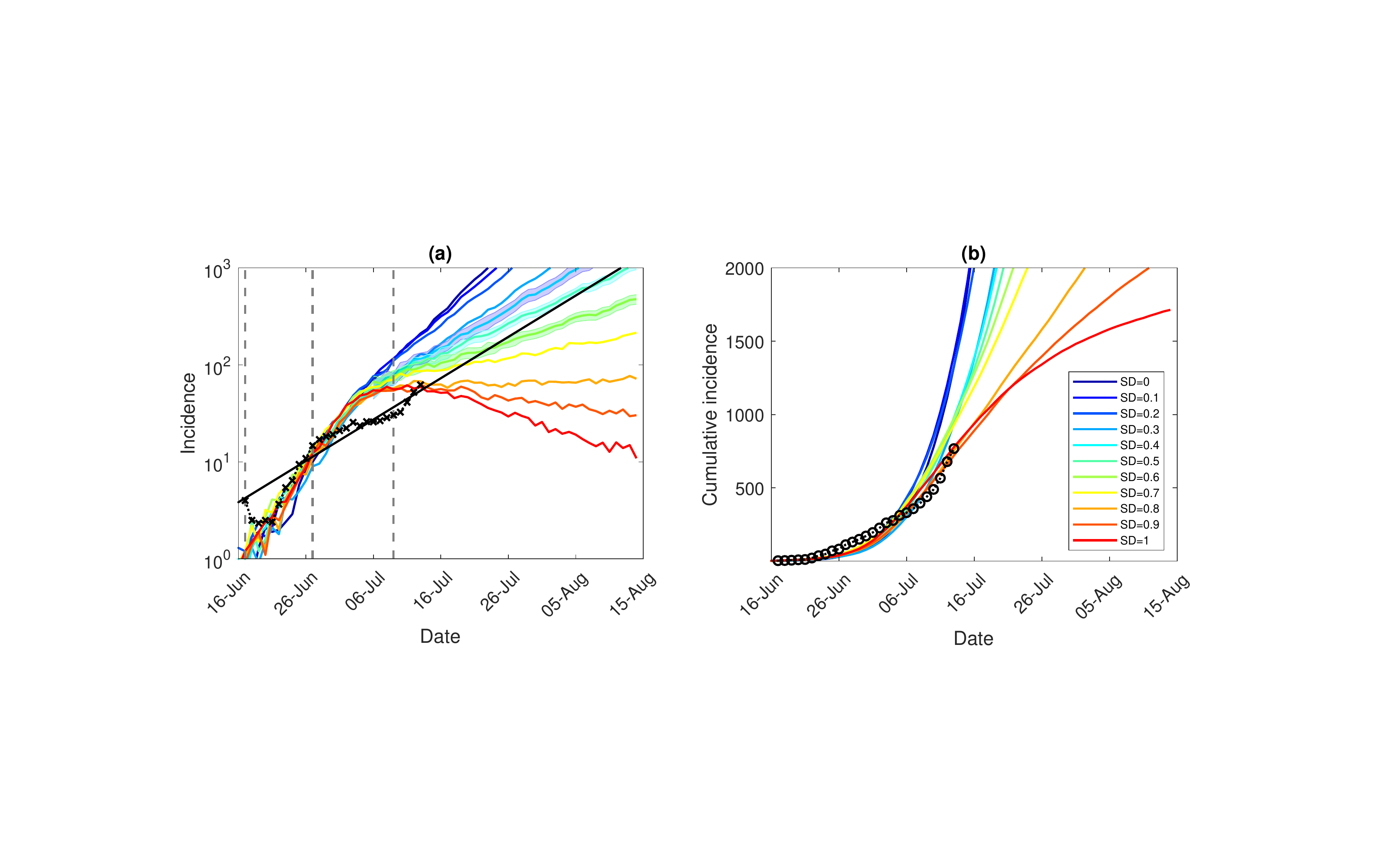}
    \caption{\textbf{ Sensitivity analysis for early interventions, moderate restrictions and shorter incubation period (NSW; progressive vaccination rollout; suppression threshold: 100 cases; $T_{inc} = 4.4$, $R_0 = 5.97$, $T_{gen} = 6.88$)}: a comparison between simulation scenarios and actual epidemic curves up to 13 July, under moderate interaction strengths ($CI_c = CI_w = 0.25$, $HQ_c = HQ_w = 0.25$, $SD_c = 0.25$, $SC = 0.5$). A moving average  of the actual time series for (a) (log-scale) incidence (crosses), and (b) cumulative incidence (circles); with an exponential fit of the incidence's moving average (black solid line showing $\beta_I = 0.098$). Vertical dashed marks align the simulated days with the outbreak start (17 June, day 11), initial restrictions (27 June, day 21), and tighter lockdown (9 July, day 33).  Traces corresponding to each social distancing (SD) compliance level are shown as average over 10 runs (coloured profiles for SD varying in increments of 10\%, i.e., between $SD = 0.0$ and $SD = 1.0$). 95\% confidence intervals for incidence profiles, for $SD \in \{0.4, 0.5, 0.6\}$, are shown as shaded areas. Each SD intervention, coupled with school closures, begins with the start of initial restrictions, when cumulative incidence exceeds 100 cases (b: inset). The alignment between simulated days and actual dates may slightly differ across separate runs. Case isolation and home quarantine are in place from the outset. }
    \label{delta_nowcasting_4p4days}
\end{figure}

\begin{figure}
    \centering
    \includegraphics[clip, trim=4.2cm 4.6cm 4.2cm 5cm, width=\textwidth]{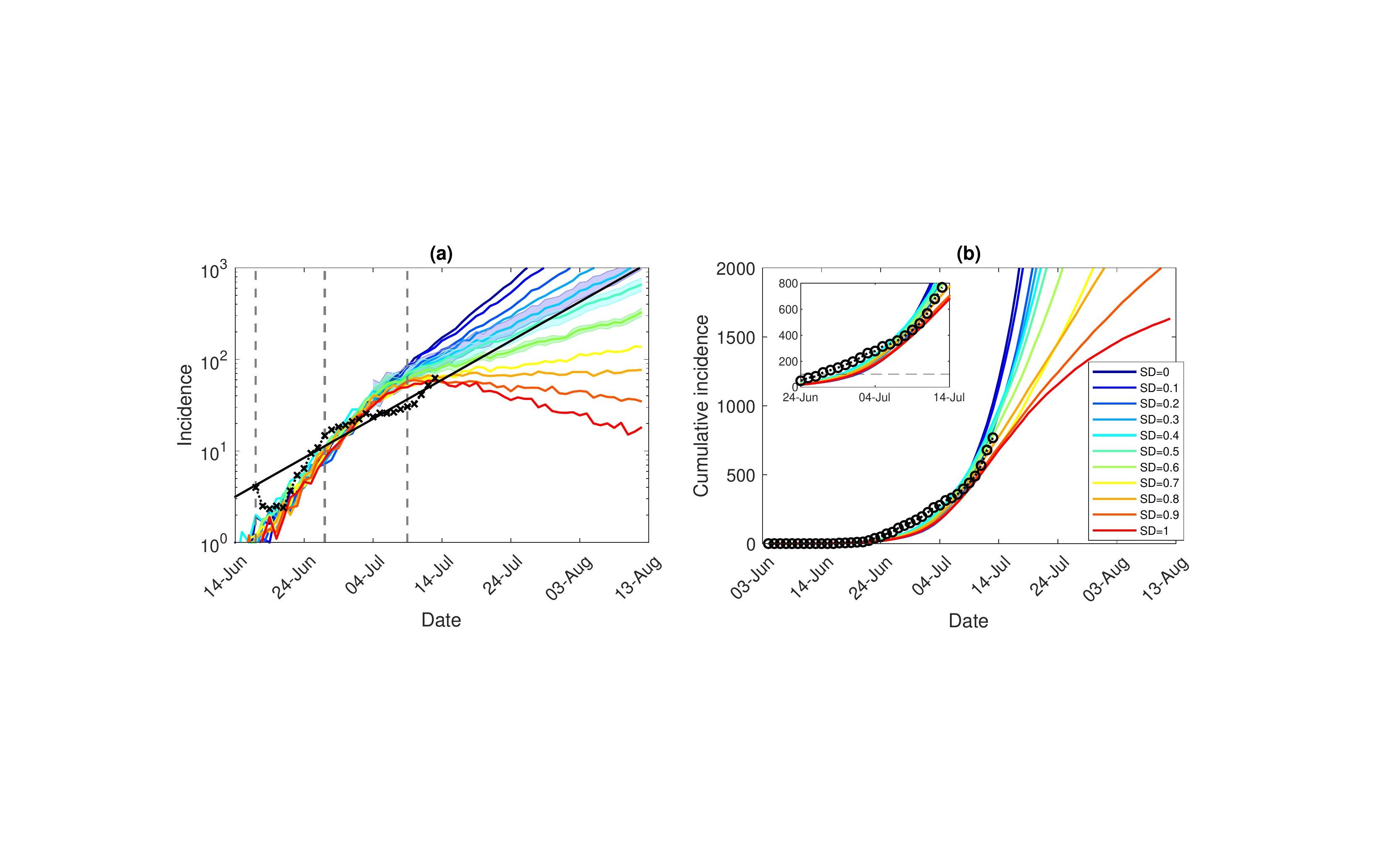}
    \caption{\textbf{Sensitivity analysis for early interventions, moderate restrictions and longer incubation period (NSW; pre-pandemic vaccination rollout; suppression threshold: 100 cases; $T_{inc} = 5.5$, $R_0 = 6.09$, $T_{gen} = 7.74$)}: a comparison between simulation scenarios and actual epidemic curves up to 13 July, under moderate interaction strengths ($CI_c = CI_w = 0.25$, $HQ_c = HQ_w = 0.25$, $SD_c = 0.25$, $SC = 0.5$).	A moving average  of the actual time series for (a) (log-scale) incidence (crosses), and (b) cumulative incidence (circles); with an exponential fit of the incidence's moving average (black solid line showing $\beta_I = 0.098$). Vertical dashed marks align the simulated days with the outbreak start (17 June, day 13), initial restrictions (27 June, day 23), and tighter lockdown (9 July, day 35).  Traces corresponding to each social distancing (SD) compliance level are shown as average over 10 runs (coloured profiles for SD varying in increments of 10\%, i.e., between $SD = 0.0$ and $SD = 1.0$). 95\% confidence intervals for incidence profiles, for $SD \in \{0.4, 0.5, 0.6\}$, are shown as shaded areas. Each SD intervention, coupled with school closures, begins with the start of initial restrictions, when cumulative incidence exceeds 100 cases (b: inset). The alignment between simulated days and actual dates may slightly differ across separate runs. Case isolation and home quarantine are in place from the outset. }
    \label{oldfig1}
\end{figure}

Furthermore, using the suppression threshold of 100 cases, corresponding to the initial restrictions (June 27), we contrasted the scenarios based on different incubation periods.  In doing so, we also varied global scalars $\kappa$ producing different reproductive numbers and generation periods, thus extending the sensitivity analysis beyond local sensitivities. Specifically, for $T_{inc} = 5.5$, the scaling factor $\kappa = 5.0$ produced the reproductive number $R_0 = 6.09$ with 95\% CI of 6.03--6.15 ($N = 6703$), and the generation period $T_{gen} = 7.74$ with 95\% CI of 7.68--7.81. For each setting, we identified the levels of social distancing (SD), triggered by the suppression threshold of 100 cases (June 27), that best matched the actual incidence data. This comparison allowed us to establish robustness of the model outcomes to changes in $T_{inc}$, $R_0$ and  $T_{gen}$. The outcomes are shown in Fig.~\ref{delta_nowcasting_4p4days} ($T_{inc} = 4.4$ and $R_0 = 5.97$, $T_{gen} = 6.88$, produced by $\kappa = 5.3$) and Fig.~\ref{oldfig1} ($T_{inc} = 5.5$ and $R_0 = 6.09$, $T_{gen} = 7.74$, produced by $\kappa = 5.0$). 

The SD levels were based on moderately reduced interaction strengths detailed in Table~1. For the setting with shorter incubation period, the best matching  scenarios  were given by $SD = 0.4$ and $SD = 0.5$, see Fig.~\ref{delta_nowcasting_4p4days}, with growth rate $\beta_{0.4} = 0.093$ being the closest match to the actual growth rate $\beta_I = 0.098$ (see Table~\ref{tab2}). For the setting with longer incubation period, the best matching  scenarios were produced by $SD = 0.3$ and $SD = 0.4$, see Fig.~\ref{oldfig1}, with $\beta_{0.3} = 0.099$ being the closest match to $\beta_I$, while  $\beta_{0.4} = 0.084$ was within the range. 
The sensitivity analysis revealed that the  model outcomes for moderate restrictions are not strongly influenced by changes in $T_{inc}$, $R_0$ and  $T_{gen}$ within the explored ranges. In other words, it confirmed the conclusion that the social distancing compliance, at least until July 13, has been followed only moderately (around $SD = 0.4$), and would be inadequate to suppress the outbreak.

\subsection*{Sensitivity of suppression outcomes for tight restrictions (counter-factual analysis)} 

\begin{figure}
    \centering
    \includegraphics[clip, trim=4.2cm 4.6cm 4.2cm 4cm, width=\textwidth]{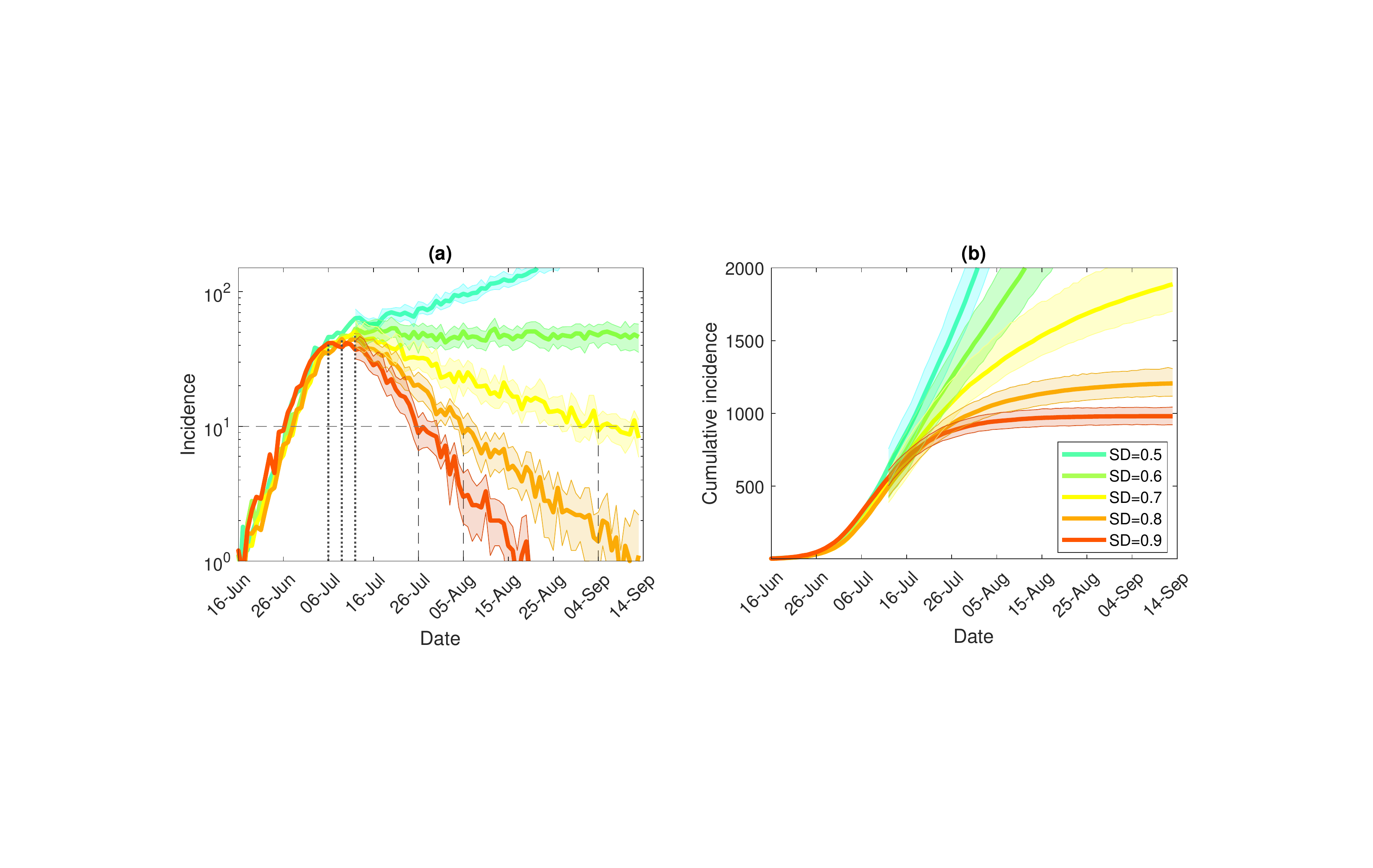}
    \caption{\textbf{Sensitivity analysis for early interventions, tight restrictions and shorter incubation period (NSW; pre-pandemic vaccination rollout; suppression threshold: 100 cases; $T_{inc} = 4.4$, $R_0 = 5.97$, $T_{gen} = 6.88$)}: counter-factual simulation scenarios, under lowest feasible  interaction strengths ($CI_c = CI_w = 0.1$, $HQ_c = HQ_w = 0.1$, $SD_c = 0.1$, $SC = 0.1$), for (a) (log scale) incidence (crosses), and (b) cumulative incidence (circles).	Traces corresponding to feasible social distancing (SD) compliance level are shown as average over 10 runs (coloured profiles for SD varying in increments of 10\%, i.e., between $SD = 0.5$ and $SD = 0.9$). Vertical lines mark the incidence peaks (dotted) and reductions below 10 daily cases (dashed). 95\% confidence intervals are shown as shaded areas. Each SD intervention, coupled with school closures, begins with the start of initial restriction, when cumulative incidence exceeds 100 cases (i.e., simulated day 21). The alignment between simulated days and actual dates may slightly differ across separate runs. Case isolation and home quarantine are in place from the outset. }
    \label{delta_suppression_4p4days}
\end{figure}

\begin{figure}
    \centering
    \includegraphics[clip, trim=4.2cm 4.6cm 4.2cm 4cm, width=\textwidth]{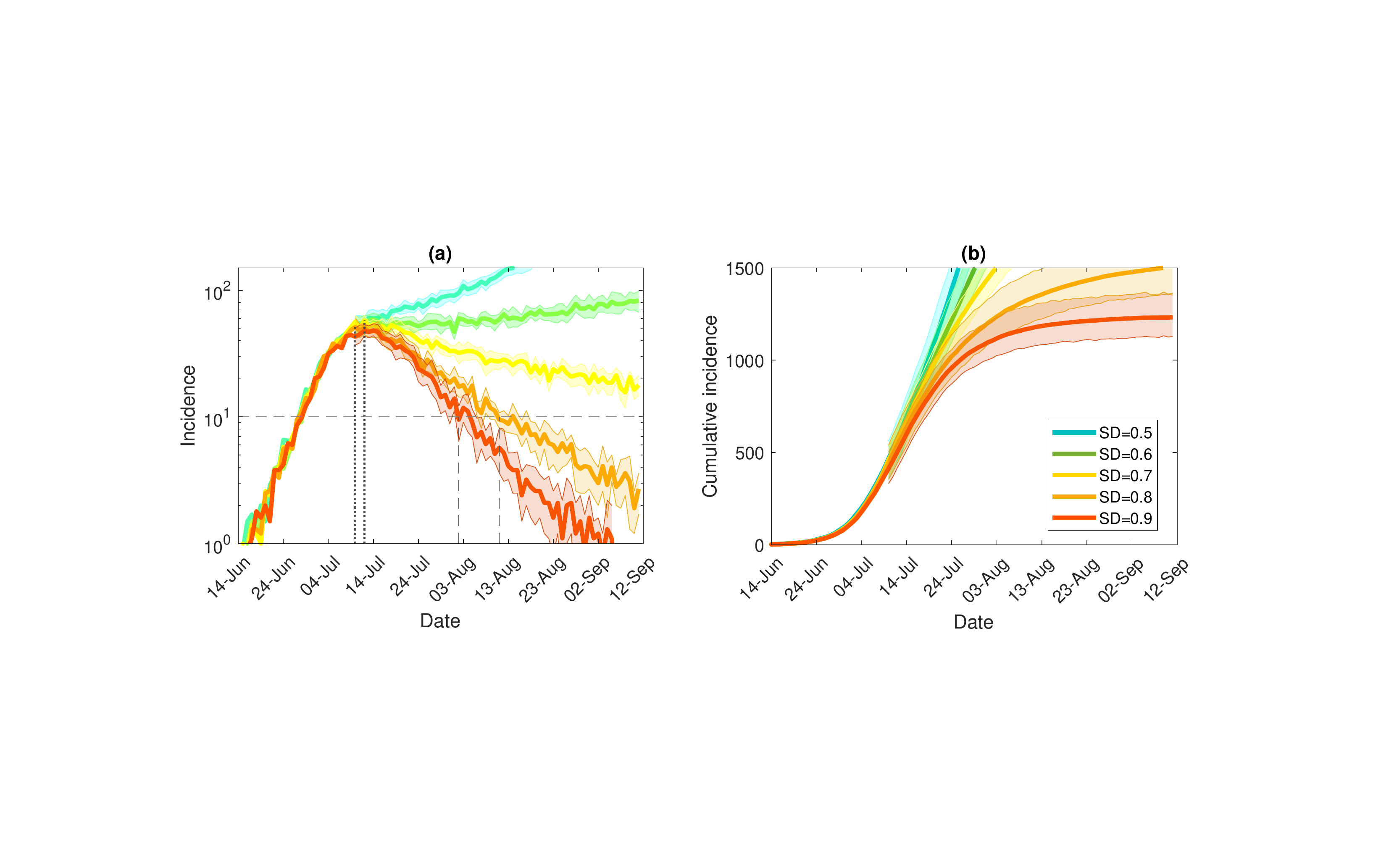}
    \caption{\textbf{Sensitivity analysis for early interventions, tight restrictions and longer incubation period (NSW; pre-pandemic vaccination rollout; suppression threshold: 100 cases; $T_{inc} = 5.5$, $R_0 = 6.09$, $T_{gen} = 7.74$)}: counter-factual simulation scenarios, under lowest feasible  interaction strengths ($CI_c = CI_w = 0.1$, $HQ_c = HQ_w = 0.1$, $SD_c = 0.1$, $SC = 0.1$), for (a) (log scale) incidence (crosses), and (b) cumulative incidence (circles). Traces corresponding to feasible social distancing (SD) compliance level are shown as average over 10 runs (coloured profiles for SD varying in increments of 10\%, i.e., between $SD = 0.5$ and $SD = 0.9$). 95\% confidence intervals are shown as shaded areas. Each SD intervention, coupled with school closures, begins with the start of initial restrictions, when cumulative incidence exceeds 100 cases (i.e., simulated day 23). The alignment between simulated days and actual dates may slightly differ across separate runs. Case isolation and home quarantine are in place from the outset. }
    \label{oldfig2}
\end{figure}

We also contrasted the suppression scenarios based on different incubation periods, reproductive numbers and generation periods (again using the threshold of 100 cases, triggered by the tight restrictions, 
under a pre-pandemic vaccination coverage). We explored feasible SD levels, $0.5 \leq SD \leq 0.9$, staying with $CI = 0.7$ and $HQ = 0.5$, but using  the lowest feasible interaction strengths ($NPI_c = 0.1$, where NPI is one of CI, HQ, SC and SD), as specified in Table~1. 
 For each setting, we identified the duration of  measures required to reduce the incidence below 10. The results are shown in Fig.~\ref{delta_suppression_4p4days} ($T_{inc} = 4.4$, $R_0 = 5.97$, $T_{gen} = 6.88$, $\kappa = 5.3$) and Fig.~\ref{oldfig2} ($T_{inc} = 5.5$, $R_0 = 6.09$, $T_{gen} = 7.74$, $\kappa = 5.0$). 

For each setting, a suppression of the outbreak is observed only for macro-distancing at $SD \geq 0.7$. Specifically, at $SD = 0.8$, new cases reduce below 10 per day approximately a month after a peak in incidence (when $T_{inc} = 5.5$, $R_0 = 6.09$), and the alternate setting ($T_{inc} = 4.4$, $R_0 = 5.97$) achieves this target a few days earlier (in 28 days). At $SD = 0.7$ the difference between the settings grows: while for the setting with $T_{inc} = 5.5$, $R_0 = 6.09$ 
the post-peak suppression period exceeds two months, the alternative ($T_{inc} = 4.4$, $R_0 = 5.97$) approaches the target about eight weeks (55 days) after the peak. There is a minor difference between the considered settings at $SD = 0.9$ which would achieve the required reduction within three weeks following the peak in incidence. 
The sensitivity analysis shows that changes in $T_{inc}$, $R_0$ and $T_{gen}$ within the considered ranges do not strongly affect the modelled suppression outcomes. This supports the projection that the peak in incidence would be followed by approximately four weeks at $SD = 0.8$, and that this period would lengthen at least twice if the compliance reduced by 10\% to $SD = 0.7$ (this setting produced the highest sensitivity among the levels demonstrating the suppression, due to its low rate of the incidence decline).

\subsection*{Fraction of symptomatic cases in children}
 
During the initial outbreak in Sydney, several COVID-19 cases have been reported in schools and early childhood centres, with one outbreak in a primary school at the end of June involving four children~\citep{danchin2021kids,nsw-jun28-2021}. On 8 September 2021, the National Centre for Immunisation Research and Surveillance (NCIRS) released a comprehensive report reviewing SARS-CoV-2 transmission within schools and early childhood services in NSW~\citep{ncirs-sep-2021}. The NCIRS report analysed transmissions of the Delta variant during the period between 16 June 2021 and 31 July 2021, with follow-up data until 19 August 2021. It noted that the majority of children (98\%) had asymptomatic or mild infection, without further separating these two categories. Importantly, between 16 June and 31 July, 22\% cases were recorded in children and young people under the age of 18 years (for the period between 1 and 19 August, this proportion was reported as 29\%). The average fraction of cases in children, $A_c = 0.27$, within the range of [0.22, 0.29], is higher than the one reported in 2020 for the ancestral lineage of SARS-CoV-2 (with the corresponding proportion reported to be as low as 3.2\%)~\citep{macartney2020transmission}, suggesting a higher fraction of symptomatic cases for children (including mild cases).  

We analysed the model sensitivity to changes in the fraction of symptomatic cases for children, varying it from $\sigma_c = 0.134$ to a higher value ($\sigma_c = 0.268$). This analysis was carried out for the incubation period $T_{inc} = 4.4$ days and the scalar $\kappa = 5.3$. 

The reproductive number increased from $R_0 = 5.97$ for the lower fraction (95\% CI of 5.93--6.00, over $N = 6318$ simulations), to $R_0 = 6.20$ for the higher fraction (with 95\% CI of 6.16--6.23, $N = 6609$). The change was within 4\%. The generation period has changed even less: from $T_{gen} = 6.88$ for the lower fraction (95\% CI of 6.81--6.94, $N = 6318$), to $T_{gen} = 6.93$ (with 95\% CI of 6.87--6.99, $N = 6609$). This change stayed within 1\%, with confidence intervals overlapping.  Such low sensitivity is in agreement with our prior analysis showing slow linear increases of $R_0$ and $T_{gen}$ in response to changes in the fraction $\sigma_c$~\citep{chang2020modelling}.

Our Australia-wide modelling used the higher fraction: $\sigma_c = 0.268$, resulting in $R_0 = 6.20$ and $T_{gen} = 6.93$. In concordance with the NCIRS report, this setting resulted in the fraction of cases in children averaging $A_c = 0.22$ (computed for the period from 2 July to 19 August, with $SD = 0.5$), and varying in the range between 0.17 and 0.31, as shown in Fig.~\ref{fig-children}. The levels $SD = 0.4$ and $SD = 0.6$ produced similar, largely overlapping profiles (see Supplementary Data 1).

\section*{Growth rates}

To estimate growth rates $\beta$, we fit a 7-day moving average of the corresponding incidence time series $I(t)$ to an exponential function $\alpha \exp(\beta (t))$, using MATLAB R2020a function $movmean(I, [6 \ 0])$. 
The growth rate of the observed incidence were estimated for several time periods (see Table~\ref{tab2}).

The growth rates $\beta_{SD}$  for the time series simulated for each SD level between 0.0 and 1.0 were estimated for the periods lasting from either the start of initial restrictions (27 June), or from 16 July (comprehensive lockdown measures were announced on 17 July).

\section*{Hospitalisations and fatalities}

In computing daily hospitalisations, as fractions of case incidence, we used age-dependent case hospitalisation risks (CHRs) reported for the Alpha (B.1.1.7) variant~\citep{nyberg2021risk}, and scaled up to the Delta variant. To scale the CHRs, we run a linear regression between (i) the hospitalisations computed according to the CHRs determined for the Alpha variant~\citep{nyberg2021risk}, and (ii) the actual hospitalisations reported in Australia between 16 June and 24 September 2021~\citep{covid19data}. The regression produced a good fit ($R^2 = 0.988$), with the multiplier of 1.715 and the additive constant of 43.38, see Fig.~\ref{fig-regr}. The CHR estimates are shown in Table~\ref{tab-CHR}. Using the case incidence projections (for vaccinated and unvaccinated agents) and the CHRs, the hospitalisations were computed with time offset of 5 days~\citep{nsw-34-2021}, and the average hospital stay of 14 days. The vaccine efficacy against severe disease was assumed to be above 90\% (specifically, \VEh = 0.95)~\citep{nasreen2021effectiveness,stowe2021effectiveness}, reducing the number of hospitalised vaccinated agents accordingly, before computing the resultant occupancy.

Age-dependent rate estimates for daily ICU admissions, as fractions of daily hospitalisations, were approximated by determining the corresponding ratios between the average ICU occupancy and the average hospitalisations (occupancy), reported in NSW between 16 June and 28 August 2021~\citep{nsw-34-2021}. These estimates are included in Table~\ref{tab-CHR}. Using the hospitalisations (for vaccinated and unvaccinated agents) and the ICU admission rates, the ICU demand was computed with the average hospital stay of 18 days~\citep{burrell2021outcomes}.
 
To estimate potential fatalities, as fractions of case incidence, we use a meta-regression equation for age-dependent infection fatality rates (IFRs)~\citep{levin2020assessing}:
\begin{equation}
    \log_{10}(\text{IFR}) = -3.27 + 0.0524 \times \text{age} 
		\label{eq-ifr}
\end{equation}
The IFR for the age over 80 years was truncated at the IFR for 80 years. Using the case incidence projections (for vaccinated and unvaccinated agents) and the IFRs, the fatalities were computed with time offset of 11 days~\citep{nsw-34-2021}. The vaccine efficacy against death was assumed to be above 90\% (\VEf = 0.95)~\citep{nasreen2021effectiveness,stowe2021effectiveness,hyde2021vaccination},  reducing the number of fatalities among vaccinated agents accordingly.

\newpage

\section*{Supplementary Tables and Figures}

\bgroup
\def\arraystretch{1.3}
\setlength\tabcolsep{4mm}
\begin{table}[h]
\centering
 \noindent
 \caption{Daily transmission rates $q_{j\to i}^g$ for different contact groups $g$. The age is assigned an integer value.}
 \label{tab:transmission-rates}
 \vspace{1mm}
 \resizebox{0.9\textwidth}{!}{%
 	{\raggedright
\begin{tabular}{l|l|l|l}
Contact Group $g$ & Infected Individual $j$ & Susceptible Individual $i$ & Transmission Probability $q^g_{j \to i}$ \\
\hline \hline
Household size 2 & Any & Child ($\le 18$) & 0.0933 \\
& Any & Adult ($\ge 19$) & 0.0393 \\
\hline
Household size 3 & Any &  Child ($\le 18$) & 0.0586 \\
 & Any & Adult ($\ge 19$) & 0.0244 \\
\hline
Household size 4 & Any & Child ($\le 18$) & 0.0417 \\
& Any & Adult ($\ge 19$) & 0.0173 \\
\hline
Household size 5 & Any & Child ($\le 18$) & 0.0321 \\
 & Any & Adult ($\ge 19$) & 0.0133 \\
\hline
Household size 6 & Any & Child ($\le 18$) & 0.0259 \\
& Any &  Adult ($\ge 19$) & 0.0107 \\
\hline
School & Child ($\le 18$) & Child ($\le 18$) & 0.000292 \\
Grade & Child ($\le 18$) & Child ($\le 18$) & 0.00158 \\
Class & Child ($\le 18$) & Child ($\le 18$) & 0.035 \\
\hline
\end{tabular}
}}
\label{transmission_table}
\end{table}
\egroup

\bgroup
\def\arraystretch{1.3}
\setlength\tabcolsep{4mm}
\begin{table}[h]
	\caption{Daily contact rates $c_{j \to i}^g$ for different contact groups $g$. The age is assigned an integer value.}
	\label{tab:contact-table}
	\vspace{1mm}
	\centering
\resizebox{0.9\textwidth}{!}{%
	{\raggedright
	 \noindent
	\begin{tabular}{l|l|l|l}
	Mixing group $g$ & Infected individual $j$ & Susceptible individual $i$ & Contact probability $c_{j \to i}^g$ \\
	\hline \hline
	Household cluster & Child ($\le 18$) & Child ($\le 18$) &  0.05 \\
	 & Child ($\le 18$) & Adult ($\ge 19$) & 0.05 \\
	 & Adult ($\ge 19$) & Child ($\le 18$) & 0.05  \\
	 & Adult ($\ge 19$) & Adult ($\ge 19$) & 0.05 \\
	\hline
	Working Group & Adult (19-64) & Adult (19-64) & 0.05 \\
	\hline
	Neighbourhood & Any & Child (0-4) &  0.0000435 \\
	 & Any & Child (5-18) & 0.0001305 \\
	 & Any & Adult (19-64) & 0.000348  \\
	 & Any & Adult ($\ge 65$) & 0.000696 \\
	\hline
	Community & Any & Child (0-4) &  0.0000109 \\
	 & Any & Child (5-18) & 0.0000326 \\
	 & Any & Adult (19-64) & 0.000087  \\
	 & Any & Adult ($\ge 65$) & 0.000174 \\
	\hline
	\end{tabular}
	}
}
\end{table}
\egroup

\bgroup 
\def\arraystretch{1.3}
\begin{table}[h]
    \caption{Main parameters for AMTraC-19 transmission model. }
    \label{tab-param}
	\vspace{1mm}
	\centering
\resizebox{1.0\textwidth}{!}{%
	{\raggedright
	 \noindent
    \begin{tabular}{c|c|c|c}
        parameter & value & distribution & notes \\
        \hline \hline
         $\kappa$ & 5.3 & NA & global transmission scalar \\
         $T_{inc}$ & 4.4 days (mean)  & lognormal ($\mu = 1.396$, $\sigma = 0.413$)  & incubation period \\
         $T_{rec}$ & 10.5 days (mean)  & uniform [7, 14] days & symptomatic (or asymptomatic) period \\
         $\alpha$ & 0.5  & NA  & asymptomatic transmission scalar \\
         $\rho$ & 0.08  & NA  & contact-to-transmission scalar \\
         $\sigma_a$ & 0.67  & NA & probability of symptoms in adults (age $>$ 18)  \\ 
         $\sigma_c$ & 0.134 or 0.268 & NA & probability of symptoms in children (age $\le$ 18)  \\ 
         $r_{symp}$ & 0.227  & NA & daily case detection rate (symptomatic)  \\ 
         $r_{asymp}$ & 0.01  & NA & daily case detection rate (asymptomatic)  \\ \hline
    \end{tabular}
	}
}
\end{table}
\egroup

\bgroup 
\def\arraystretch{1.3}
\begin{table}[h]
	\caption{The macro-distancing parameters and interaction strengths of NPIs. An example simulation scenario set for 28 weeks (196 days), with SD and SC synchronised to be triggered by 400 cumulative cases and last for 121 days.}
	\label{tabX}
	\vspace{1mm}
	\centering
\resizebox{1.0\textwidth}{!}{%
	{\raggedright
	 \noindent
	 \begin{tabular}{l|c|c|c|c|c|c|c}
	 & \multicolumn{3}{c|}{Macro-distancing} & \multicolumn{4}{c}{Micro-distancing (interaction strengths)}  \\ \hline
	Intervention & Compliance level & Duration $T$ & Threshold  & Household & Community & Workplace/School & Duration $t$ \\
	\hline \hline
	CI & 0.7 & 196 & 0  & 1.0 & 0.25 & 0.25 & $D(i)$ \\
  HQ & 0.5 & 196 & 0  & 2.0 & 0.25 & 0.25 & 14 \\
	SC$^c$ & 1.0 & 121 & 400  & 1.0 & 0.5 & 0 & 121 \\
	SC$^a$ & 0.5 & 121 & 400 & 1.0 & 0.5 & 0 & 121  \\
	SD & 0.5 & 121 & 400 & 1.0 & 0.25 & 0.1 & 121  \\
	\hline 
	\end{tabular}
	}
}
\end{table}
\egroup

\bgroup 
\def\arraystretch{1.3}
\begin{table}[h]
	\caption{Simulated and actual~\citep{aus--2021} vaccination coverage across Australia (double vaccinated individuals).}
	\label{tab-vacc}
	\vspace{1mm}
	\centering
\resizebox{0.9\textwidth}{!}{%
	{\raggedright
	 \noindent
	 \begin{tabular}{l|c|c|c}
	date & Adults (16+): actual (\%) & Adults (16+): simulated (\%) & Total population: simulated (\%) \\
	\hline \hline
  16 July   & 13.35 & 17.28 & 13.82 \\
  16 August   & 26.88 & 33.51 & 26.80 \\
  15 September   & 44.66 & 49.21 & 39.36 \\
  16 October   & 67.85 & 65.44 & 52.35 \\
	13 November   & 83.01 & 79.38 & 63.49 \\
	\hline 
	\end{tabular}
	}
}
\end{table}
\egroup

\bgroup 
\def\arraystretch{1.3}
\begin{table}[h]
    \caption{Calibration targets. }
    \label{tab-calibr}
	\vspace{1mm}
	\centering
\resizebox{1.0\textwidth}{!}{%
	{\raggedright
	 \noindent
    \begin{tabular}{c|c|c|c|c}
        parameter & value from ABM [range] & sample size & target value [range] & notes \\
        \hline \hline
         $R_0$ ($\sigma_c = 0.134$) & 5.97 [5.93, 6.00] & 6318 & [5.5, 6.5] & basic reproductive ratio~\citep{campbell2021increased} \\
         $T_{gen}$ ($\sigma_c = 0.134$) & 6.88 [6.81, 6.94] & 6318 & [5.8, 8.1] & generation/serial interval \citep{campbell2021increased}\\
				& & & &  \citep{wu2020estimating,blanquart2021spread} \\ \hline
								$R_0$ ($\sigma_c = 0.268$) & 6.20 [6.16, 6.23] & 6609 & [5.5, 6.5] & basic reproductive ratio \citep{campbell2021increased} \\ 
         $T_{gen}$ ($\sigma_c = 0.268$) & 6.93 [6.87, 6.99] & 6609 & [5.8, 8.1] & generation/serial interval \citep{campbell2021increased} \\
				& & & &  \citep{wu2020estimating,blanquart2021spread} \\ \hline
         $\beta_{I}$ & $\beta_{0.4} =$ 0.084 & 10 & 0.098 [0.084, 0.112]  & growth rate, case incidence (NSW: 17 June--13 July) \\
         $\beta_{III}$ & $\beta_{0.6} =$ 0.029 & 10 & 0.037 [0.026, 0.048]  & growth rate, case incidence (NSW: 16--25 July) \\ \hline
         $A_c$ ($\sigma_c = 0.268$) & $A_c \ (SD=0.5) =$ 0.22 [0.16, 0.29]  & 10 & 0.27 [0.22, 0.29]  & fraction of  cases in children (age $\le$ 18)  \\
         &  & & & (NSW: 16 June--19 August)~\citep{ncirs-sep-2021} \\
				\hline
    \end{tabular}
	}
}
\end{table}
\egroup

\bgroup 
\def\arraystretch{1.3}
\begin{table}[h]
	\caption{The growth rate of the observed incidence.}
	\label{tab2}
	\centering
\resizebox{0.6\textwidth}{!}{%
	{\raggedright
	 \noindent
	 \begin{tabular}{l|c|c|c}
	growth rate & period & mean & 95\% CI  \\
	\hline \hline
  $\beta_{I}$   & \ 17 June -- 13 July \ & \ 0.098 \ & [0.084, 0.112] \\
  $\beta_{II}$  & \ 17 June -- 25 July \ & \ 0.076 \ & [0.069, 0.084] \\
  $\beta_{III}$ & \ 16 July -- 25 July \ & \ 0.037 \ & [0.026, 0.048] \\ 
	\hline 
	\end{tabular}
	}
}
\end{table}
\egroup

\bgroup 
\def\arraystretch{1.3}
\begin{table}[h]
	\caption{Estimates of age-dependent case hospitalisation risks (CHRs, \%) and ICU admission rates.}
	\label{tab-CHR}
	\centering
\resizebox{0.9\textwidth}{!}{%
	{\raggedright
	 \noindent
	 \begin{tabular}{l|c|c|c|c|c|c|c|c|c}
	Rate \textbackslash Age & 0-9 & 10-19 & 20-29 & 30-39 & 40-49 & 50-59 & 60-69 & 70-79 & 80+ \\
	\hline \hline
  CHR (Alpha)~\citep{nyberg2021risk} & 0.9  & 0.7  & 1.9  & 3.4  & 5.0  & 7.2  &  10.6 & 16.9  & 21.7  \\
  CHR (Delta) & 1.54  & 1.20  & 3.26  & 5.83  & 8.58  & 12.35  &  18.18 & 28.98  & 37.22  \\ \hline
  ICU (Delta) & 0.01  & 0.06  & 0.09  & 0.9 & 0.13  & 0.19  & 0.19  & 0.17  & 0.09   \\
	\hline 
	\end{tabular}
	}
}
\end{table}
\egroup

\bgroup 
\def\arraystretch{1.3}
\begin{table}[h]
	\caption{Estimates (across Australia) of the peak demand in hospitalisations and ICUs, for various vaccine efficacies (\VEi = 0.5 is used in primary analysis, varied to \VEi = 0.35 and \VEi = 0.65 in sensitivity analysis).}
	\label{tab-hosp-supp}
	\centering
\resizebox{1.0\textwidth}{!}{%
	{\raggedright
	 \noindent
	 \begin{tabular}{l|c|c|c|c|c|c}
	 & \multicolumn{3}{c|}{Peak hospitalisations: mean and 95\% CI} & \multicolumn{3}{c}{Peak ICU demand: mean and 95\% CI}  \\ \hline
	Scenario & Total & Vaccinated & Unvaccinated & Total & Vaccinated & Unvaccinated  \\ 
	\hline \hline
	\VEi \ = 0.5 & & & & & & \\ 
$SD = 0.4$ & 4805 [4282, 5257] & 113 [102, 122] & 4706 [4221, 5154] & 812 [731, 885] & 18 [16, 20] & 795 [711, 868]  \\	
$SD = 0.5$ & 1604 [1358, 1844] & 33 [29, 38] & 1575 [1331, 1809] & 272 [230, 312] & 5.7 [4.8, 6.5] & 267 [227, 306]  \\	
$SD = 0.6$ & 533 [476, 579] & 8.7 [7.7, 9.3] & 526 [467, 571] & 91 [80, 99] & 1.5 [1.3, 1.6] & 89 [79, 97]  \\
	\hline \hline
 \VEi \ = 0.35 & & & & & &  \\ 
$SD = 0.4$ & 7216 [6460, 7771]	& 158 [176, 188] &	7061 [6296, 7616]	& 1225 [1093, 1314]	& 28 [25, 31]	& 1197 [1069, 1284]  \\	
$SD = 0.5$ & 2249 [1882, 2657]	& 45 [53, 61]	& 2204 [1851, 2600]	& 381 [320, 449]	& 8.6 [7.2, 10]	& 374 [314, 441]  \\	
$SD = 0.6$ & 620 [571, 668]	& 10.9 [11.8, 12.5]	& 610 [562, 658]	& 105 [97, 114]	& 2 [1.9, 2.2]	& 103 [95, 111]  \\
	\hline 
 \VEi \ = 0.65 & & & & & &  \\ 
$SD = 0.4$ & 3817 [3497, 4068] &	69.6 [63, 75] &	3747 [3428, 3990] &	650 [598, 692] &	14.1 [13, 15] &	637 [589, 681]   \\	
$SD = 0.5$ & 1252 [1172, 1336] &	22 [21, 24] &	1230 [1148, 1313] &	213 [200, 228] &	4.1 [3.8, 4.4] &	210 [195, 224]   \\	
$SD = 0.6$ & 467 [414, 525] &	6.9 [6.1, 7.7] &	462 [411, 522] &	79 [70, 89] &	1.2 [1, 1.4] &	78 [70, 88]  \\
\hline
	\end{tabular}
	} 
}
\end{table}
\egroup

\bgroup 
\def\arraystretch{1.3}
\begin{table}[h]
	\caption{Estimates (across Australia) of cumulative fatalities (15 October 2021), for various vaccine efficacies (\VEi = 0.5 is used in primary analysis, varied to \VEi = 0.35 and \VEi = 0.65 in sensitivity analysis).}
	\label{tab-deaths-supp}
	\centering
\resizebox{0.6\textwidth}{!}{%
	{\raggedright
	 \noindent
	 \begin{tabular}{l|c|c|c}
	 & \multicolumn{3}{c}{Cumulative fatalities: mean and 95\% CI} \\ \hline
	Scenario & Total & Vaccinated & Unvaccinated \\ 
	\hline \hline
		\VEi \ = 0.5 & & &  \\ 
$SD = 0.4$ & 1201 [1057, 1326] & 10 [8.8, 11.2] & 1191 [1047, 1313] \\	
$SD = 0.5$ & 539 [479, 624] & 4.2 [3.5, 4.9] & 535 [447, 615] \\	
$SD = 0.6$ & 235 [209, 256] & 1.7 [1.5, 1.8] & 233 [205, 253] \\
	\hline \hline
 \VEi \ = 0.35 & & &  \\ 
$SD = 0.4$ & 1672 [1455, 1834] &	14.3 [12.3, 15.7] &	1658 [1435, 1817] \\	
$SD = 0.5$ & 667 [553, 811] &	5.4 [4.5, 6.5] &	661 [543, 798] \\	
$SD = 0.6$ & 259 [238, 280] &	1.9 [1.8, 2.1] &	257 [237, 278] \\
	\hline 
 \VEi \ = 0.65 & & &  \\ 
$SD = 0.4$ & 1094 [997, 1175] &	8.9 [8.1, 9.7] &	1082 [987, 1165] \\	
$SD = 0.5$ & 461 [429, 494] &	3.6 [3.4, 3.8] &	458 [426, 490] \\	
$SD = 0.6$ & 216 [191, 244] &	1.5 [1.3, 1.7] &	214 [189, 241] \\
	\hline
	\end{tabular}
	} 
}
\end{table}
\egroup

\newpage

\begin{figure}
    \centering
    \includegraphics[clip, trim=1cm 8cm 1.2cm 9cm, width=0.7\textwidth]{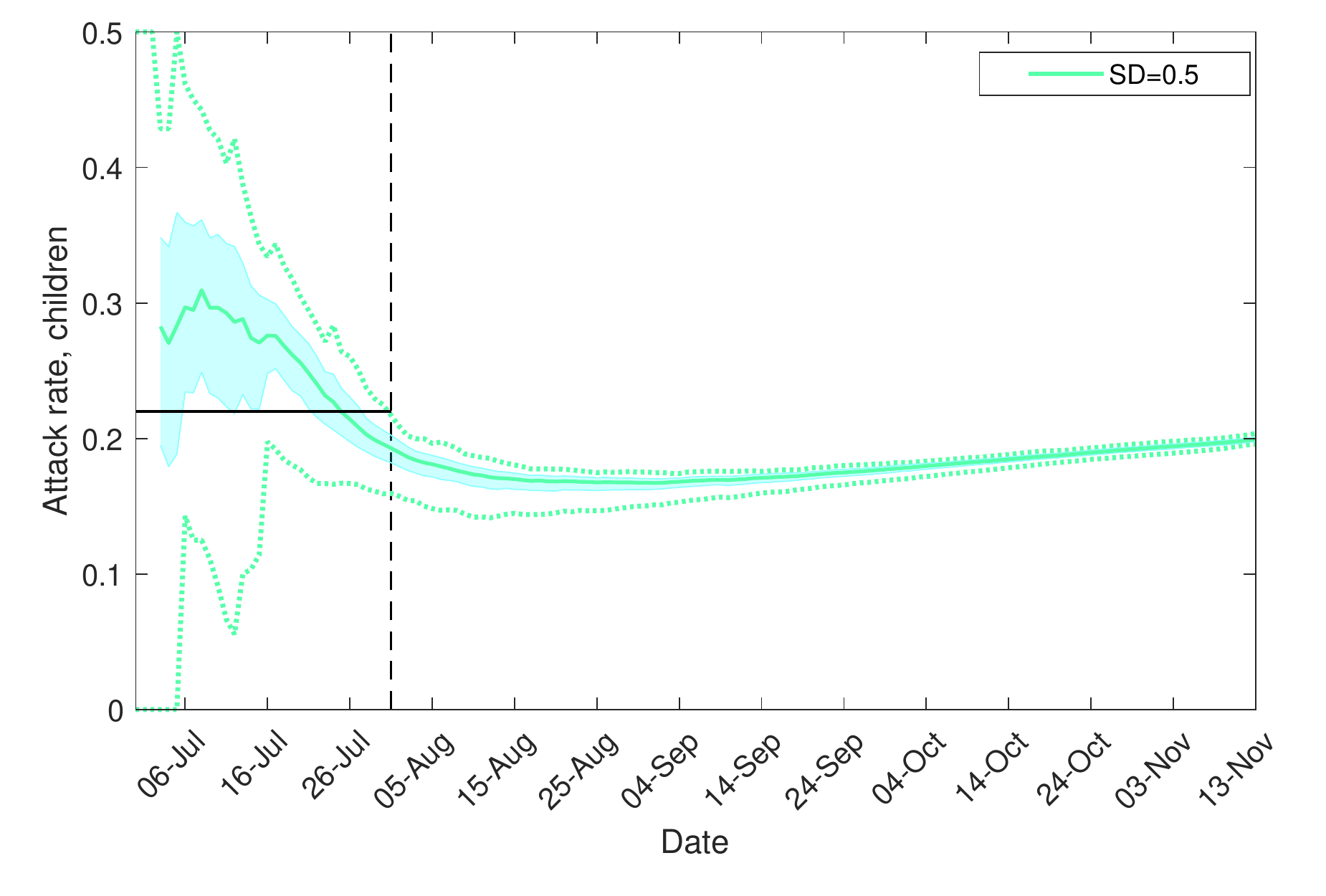}
    \caption{\textbf{Fraction of cases in children across Australia.} A trace corresponding to social distancing level $SD = 0.5$  is shown for the period between 30 June and 13 November as average over 10 runs. 95\% confidence interval is shown as a shaded area. Minimal and maximal traces, per time point, are shown with dotted lines. The horizontal line shows the fraction of cases in children averaged for the period up to 31 July ($A_c = 0.22$), in agreement with the rate reported for the period between 16 June and 31 July~\citep{ncirs-sep-2021}. The SD intervention, coupled with school closures, begins with the start of initial restrictions. The alignment between simulated days and actual dates may slightly differ across separate runs. Case isolation and home quarantine are in place from the outset. }
    \label{fig-children}
\end{figure}

\begin{figure}
    \centering
    \includegraphics[clip, trim=1cm 8.6cm 0.2cm 9cm, width=0.7\textwidth]{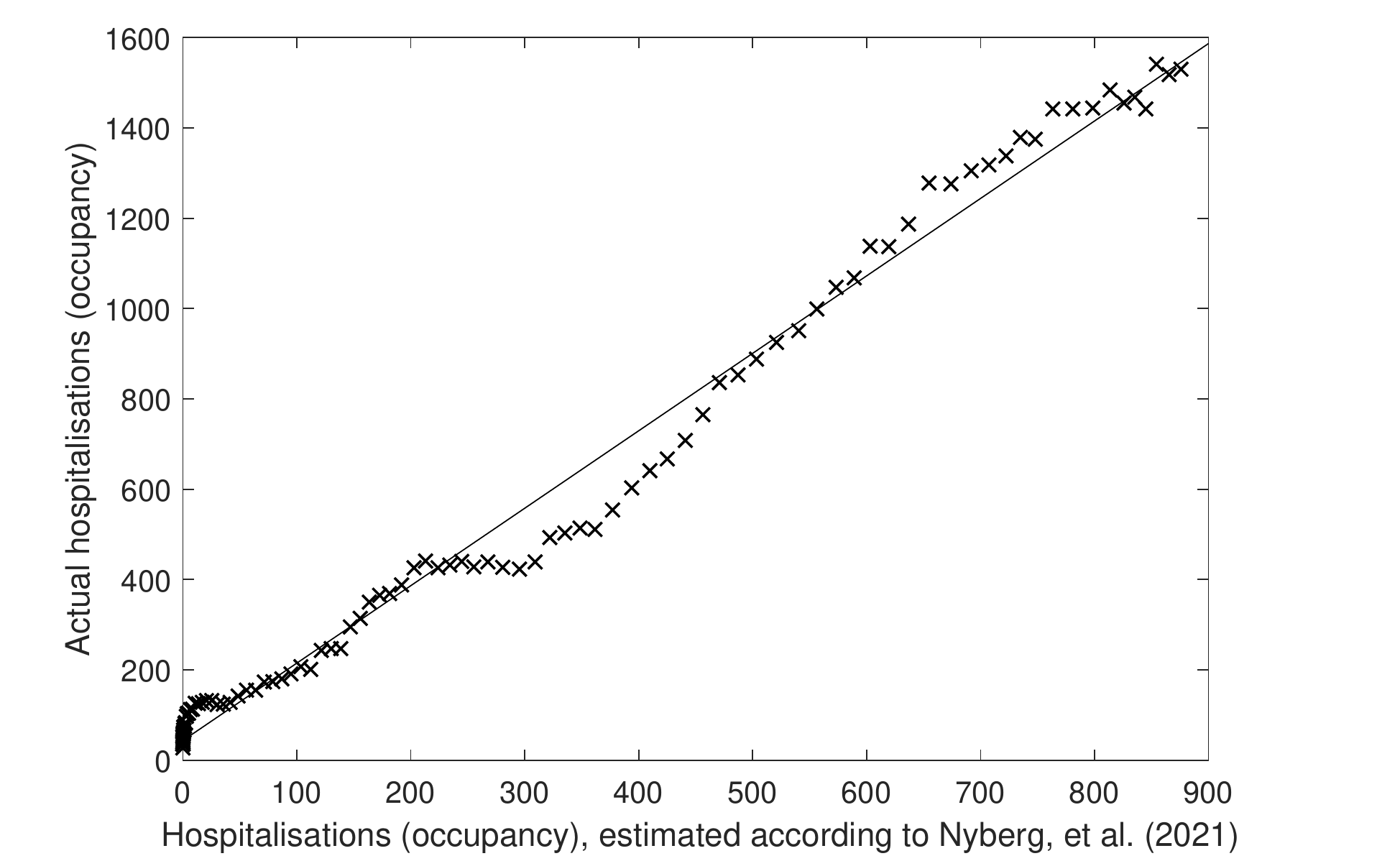}
    \caption{\textbf{Scaling case hospitalisation risks (CHRs).} A linear regression ($R^2 = 0.988$) between (i) the hospitalisations computed according to the CHRs determined for the Alpha variant~\citep{nyberg2021risk}, and (ii) the actual hospitalisations reported in Australia~\citep{covid19data} between 16 June and 24 September 2021. }
    \label{fig-regr}
\end{figure}

\begin{figure}
    \centering
    \includegraphics[clip, trim=1cm 8.5cm 1.2cm 9cm, width=0.7\textwidth]{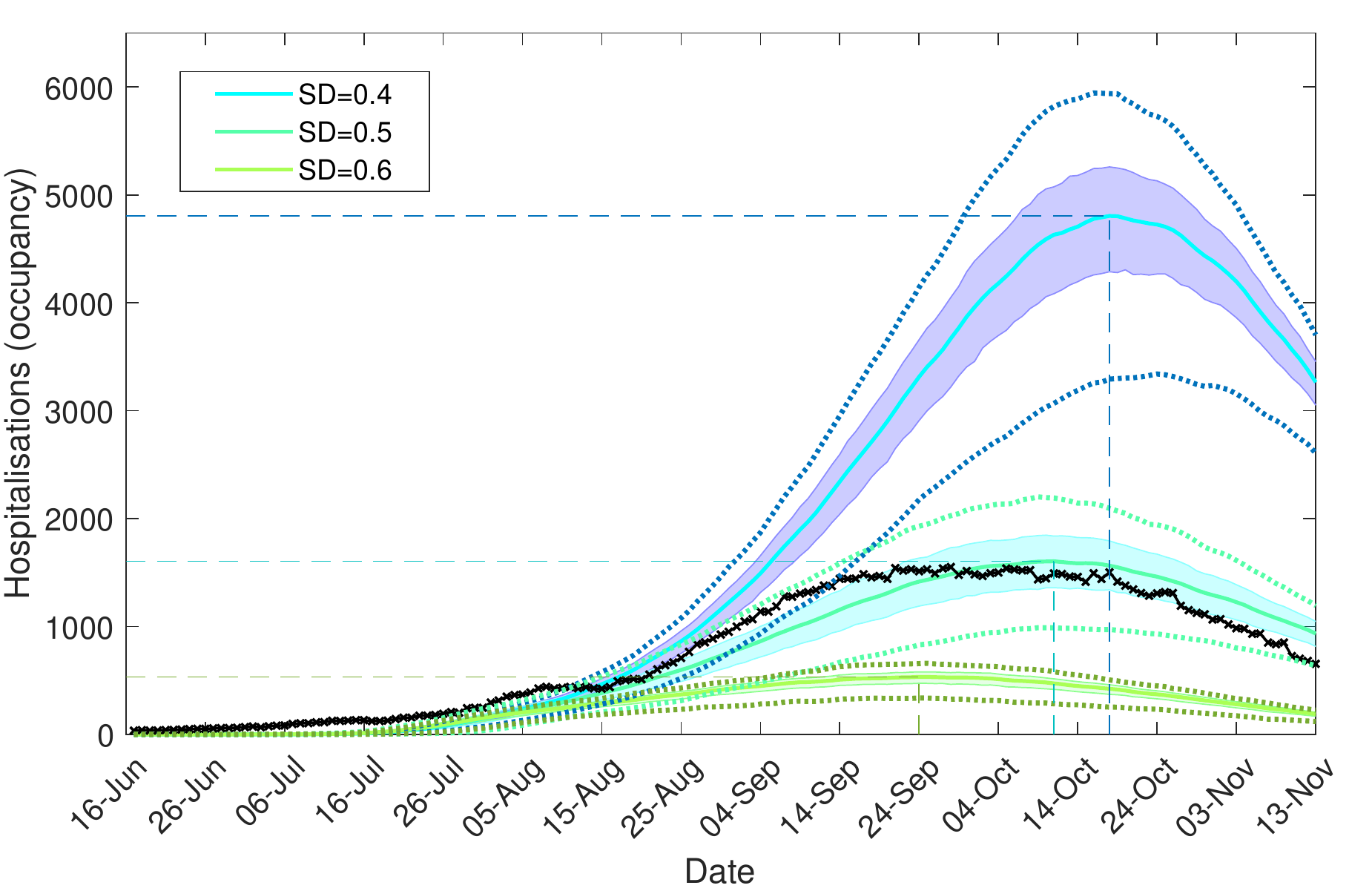}
    \caption{\textbf{Estimated hospitalisations (occupancy) across Australia.} Traces corresponding to social distancing levels $SD \in \{0.4, 0.5, 0.6\}$  are shown for the period between 16 June and 13 November, as averages over 10 runs (coloured profiles). 95\% confidence intervals are shown as shaded areas. For each SD level, minimal and maximal traces, per time point, are shown with dotted lines. A moving average  of the actual time series is shown in black. Peaks formed during the suppression period for each SD profile are identified with coloured dashed lines.  The SD intervention, coupled with school closures, begins with the start of initial restrictions. The alignment between simulated days and actual dates may slightly differ across separate runs. Case isolation and home quarantine are in place from the outset. }
    \label{fig-hosp}
\end{figure}

\begin{figure}
    \centering
    \includegraphics[clip, trim=1cm 8.5cm 1.2cm 9cm, width=0.7\textwidth]{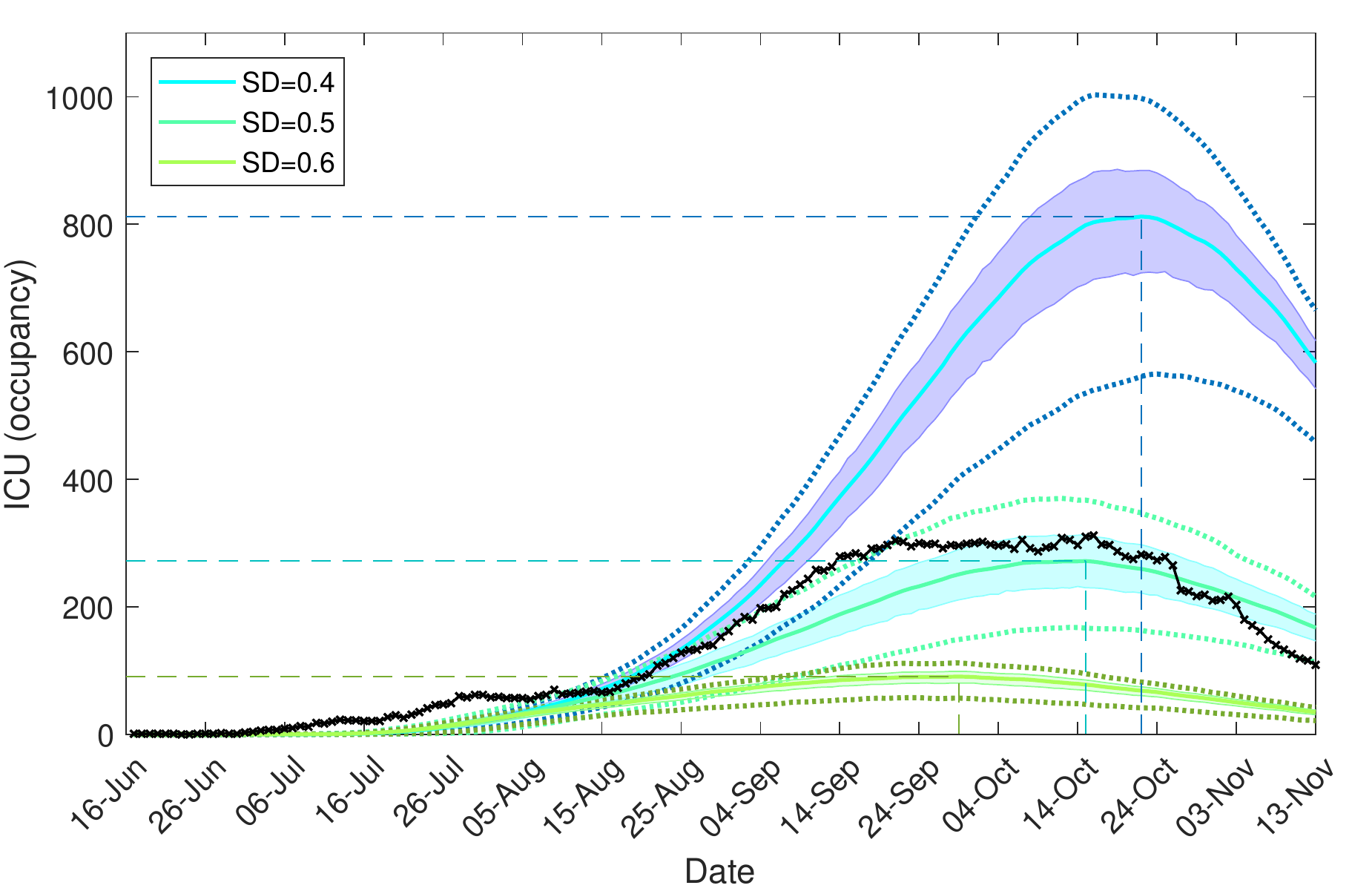}
    \caption{\textbf{Estimated ICU demand across Australia.} Traces corresponding to social distancing levels $SD \in \{0.4, 0.5, 0.6\}$  are shown for the period between 16 June and 13 November, as averages over 10 runs (coloured profiles). 95\% confidence intervals are shown as shaded areas. For each SD level, minimal and maximal traces, per time point, are shown with dotted lines. A moving average  of the actual time series is shown in black. Peaks formed during the suppression period for each SD profile are identified with coloured dashed lines. The SD intervention, coupled with school closures, begins with the start of initial restrictions. The alignment between simulated days and actual dates may slightly differ across separate runs. Case isolation and home quarantine are in place from the outset. }
    \label{fig-ICU}
\end{figure}

\begin{figure}
    \centering
    \includegraphics[clip, trim=1cm 8.5cm 1.2cm 9cm, width=0.7\textwidth]{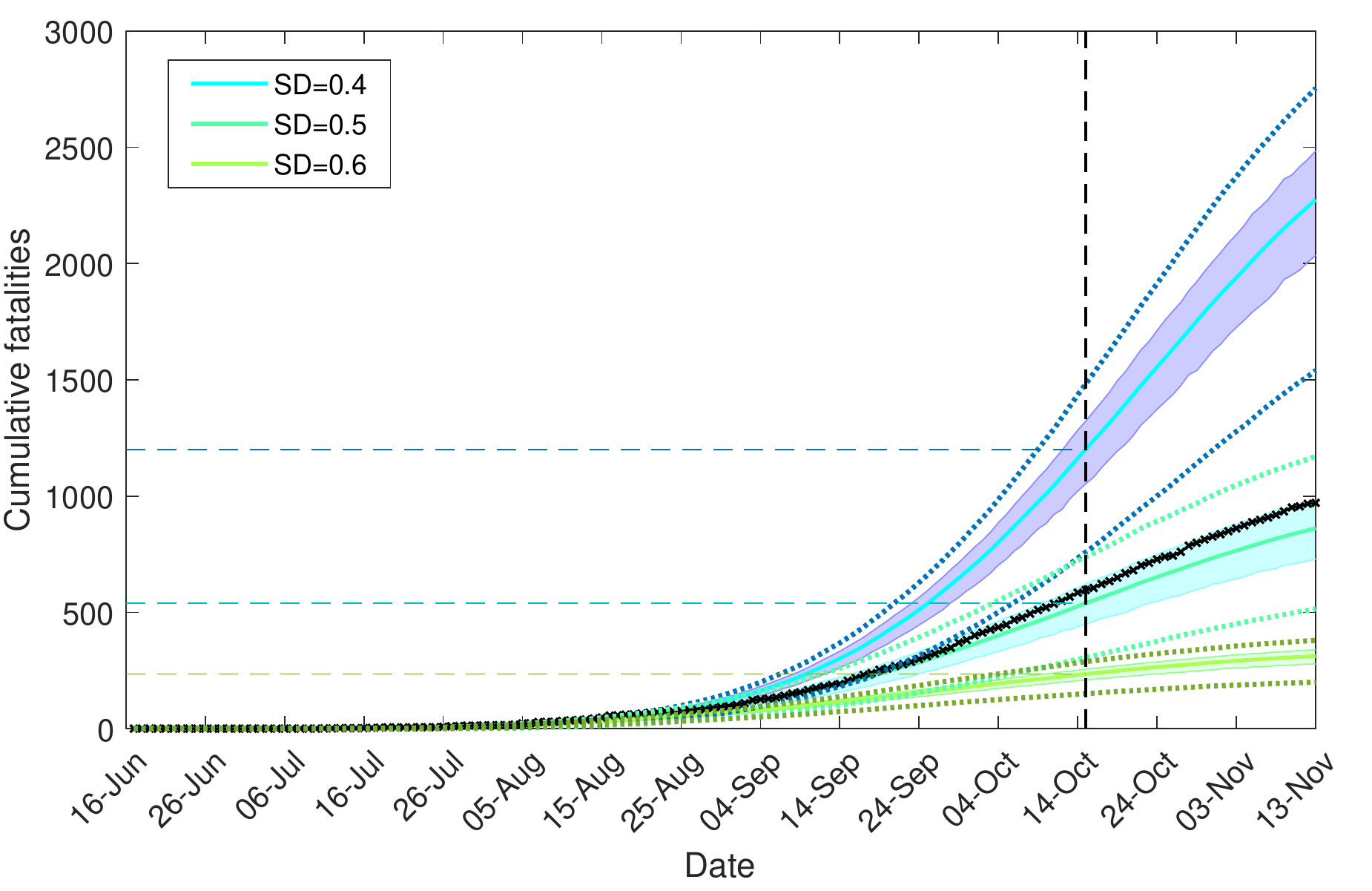}
    \caption{\textbf{Estimated cumulative fatalities across Australia.} Traces corresponding to social distancing levels $SD \in \{0.4, 0.5, 0.6\}$  are shown for the period between 16 June and 13 November, as averages over 10 runs (coloured profiles). 95\% confidence intervals are shown as shaded areas. For each SD level, minimal and maximal traces, per time point, are shown with dotted lines. A moving average  of the actual time series is shown in black. Dashed lines mark cumulative fatalities estimated for 15 October 2021, as reported in Table~3. The SD intervention, coupled with school closures, begins with the start of initial restrictions. The alignment between simulated days and actual dates may slightly differ across separate runs. Case isolation and home quarantine are in place from the outset. }
    \label{fig-deaths}
\end{figure}

\begin{figure}
    \centering
    \includegraphics[clip, trim=1cm 8.5cm 1.2cm 9cm, width=0.7\textwidth]{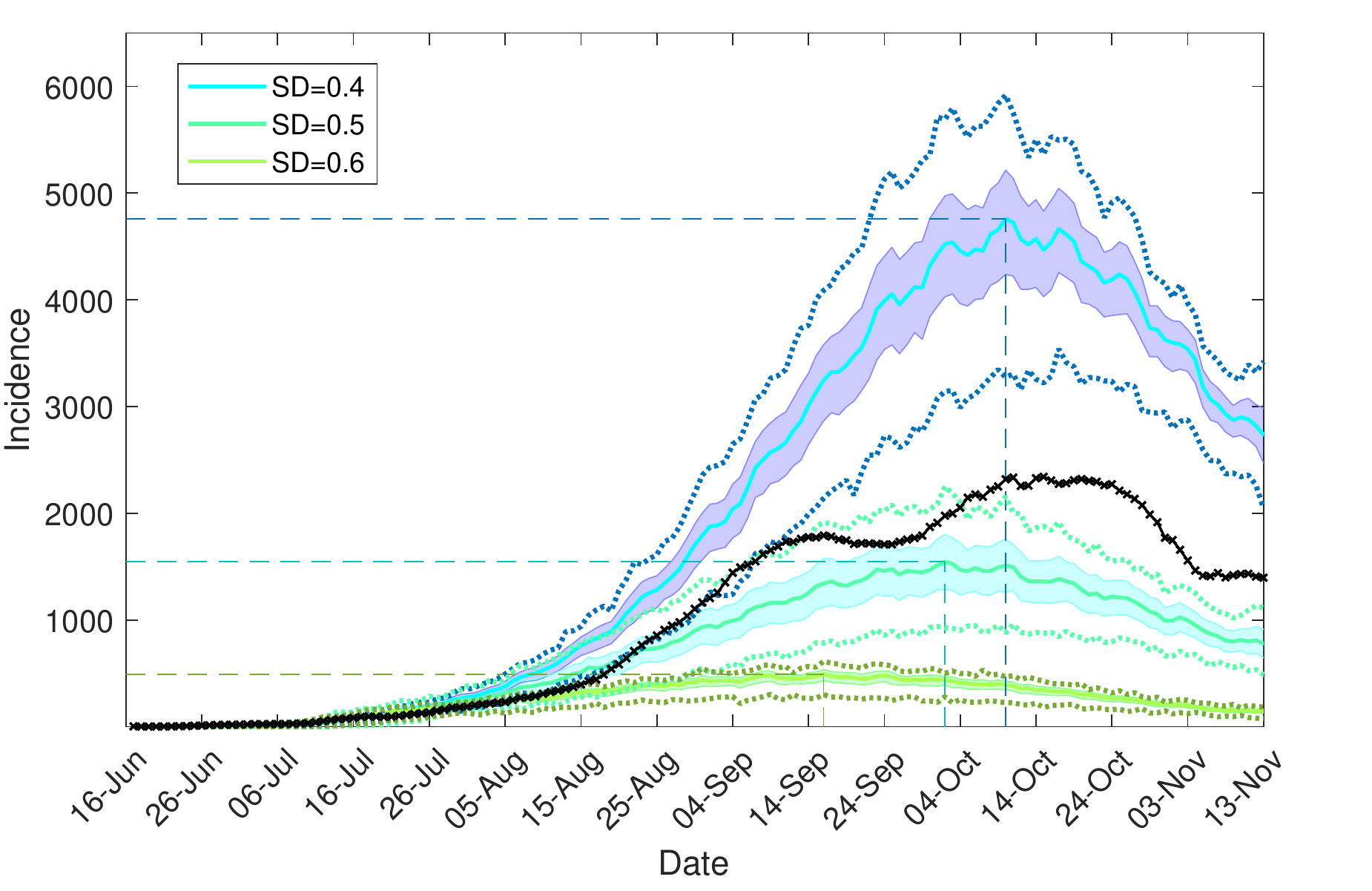}
    \caption{\textbf{Moderate restrictions (Australia; progressive vaccination rollout; suppression threshold: 400 cases)}: a comparison between simulation scenarios and actual epidemic curves up to November 13, under moderate interaction strengths ($CI_c = CI_w = 0.25$, $HQ_c = HQ_w = 0.25$, $SD_c = 0.25$, $SC = 0.5$). A moving average  of the actual time series for incidence is shown with crosses. Traces corresponding to social distancing levels $SD \in \{0.4, 0.5, 0.6\}$  are shown for the period between 16 June and 13 November, as averages over 10 runs (coloured profiles). 95\% confidence intervals are shown as shaded areas. For each SD level, minimal and maximal traces, per time point, are shown with dotted lines. Peaks formed during the suppression period for each SD profile are identified with coloured dashed lines. Each SD intervention, coupled with school closures, begins with the start of initial restrictions. The alignment between simulated days and actual dates may slightly differ across separate runs. Case isolation and home quarantine are in place from the outset. Compare with Fig.~3.a showing the same data and simulated scenarios on a log scale.}
    \label{fig3linear}
\end{figure}

\begin{figure}
    \centering
    \includegraphics[clip, trim=0.4cm 4cm 0.6cm 4cm, width=\textwidth]{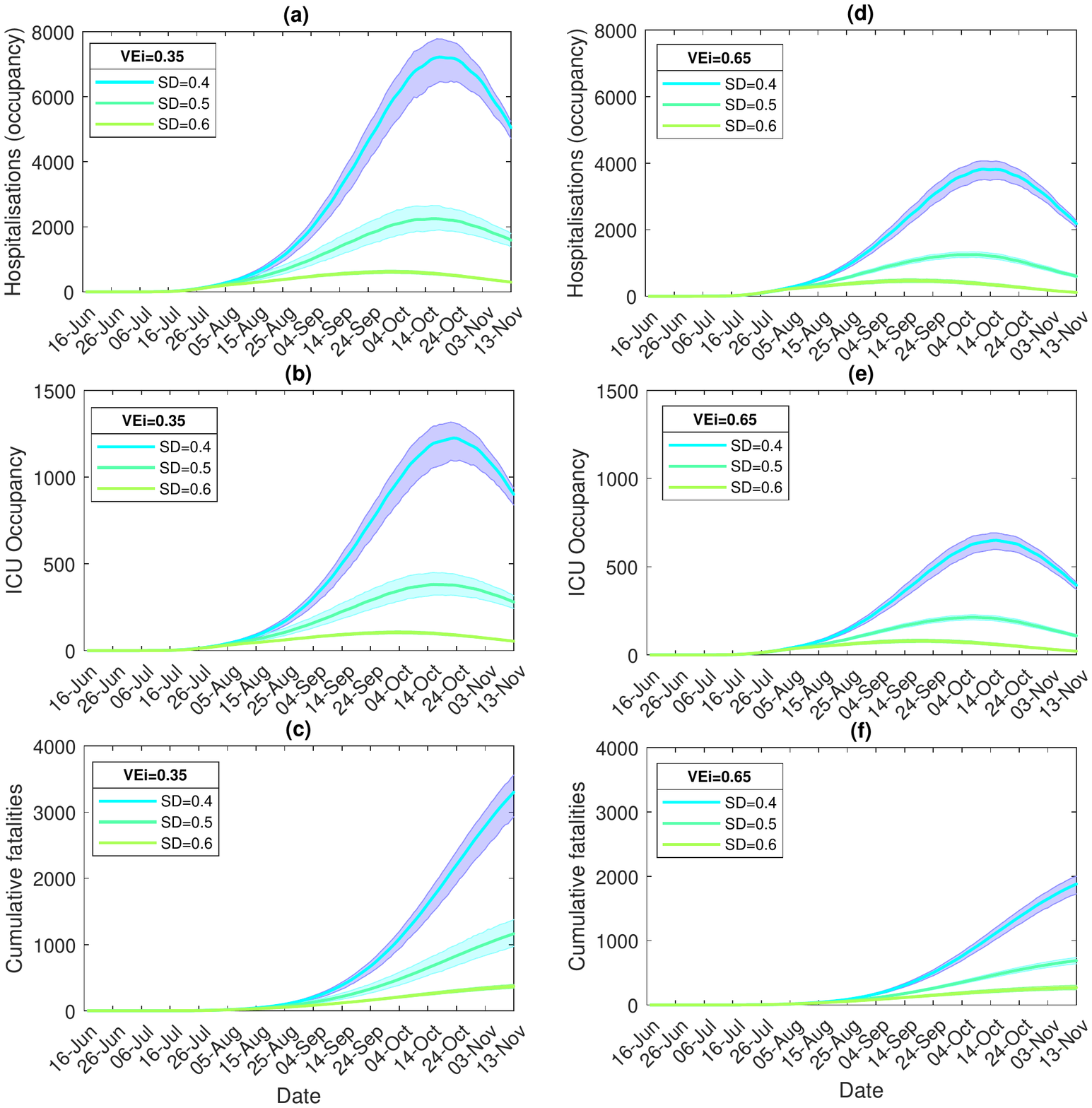}
    \caption{\textbf{Estimated hospitalisations (occupancy), ICU demand, and cumulative fatalities across Australia; sensitivity analysis for different efficacies against infectiousness}: \VEi = 0.35, shown in left panels (a), (b), (c), and \VEi = 0.65, shown in right panels (d), (e), (f). Traces corresponding to social distancing levels $SD \in \{0.4, 0.5, 0.6\}$  are shown for the period between 16 June and 13 November, as averages over 10 runs (coloured profiles). 95\% confidence intervals are shown as shaded areas. The SD intervention, coupled with school closures, begins with the start of initial restrictions. The alignment between simulated days and actual dates may slightly differ across separate runs. Case isolation and home quarantine are in place from the outset. }
    \label{figVEi-sens}
\end{figure}

\end{document}